\documentclass{pasj00}
\usepackage{graphicx}
\begin{document}
\SetRunningHead{Y. Takeda and A. Tajitsu}{Observational characteristics 
of Li-enhanced giants}
\Received{2017/04/12}
\Accepted{2017/06/06}

\title{On the observational characteristics of lithium-enhanced 
giant stars in comparison with normal red giants
\thanks{Based on data collected at Subaru Telescope,
operated by the National Astronomical Observatory of Japan.}
}

\author{
Yoichi \textsc{Takeda}\altaffilmark{1}
and
Akito \textsc{Tajitsu}\altaffilmark{2}
}
\altaffiltext{1}{National Astronomical Observatory, 2-21-1 Osawa, Mitaka, Tokyo 181-8588}
\email{takeda.yoichi@nao.ac.jp}
\altaffiltext{2}{Subaru Telescope, 650 N. A'ohoku Place, Hilo, HI 96720, U.S.A.}
\email{tajitsu@subaru.naoj.org}



\maketitle

\begin{abstract}
While lithium is generally deficient in the atmosphere of evolved giant stars 
because of the efficient mixing-induced dilution, a small fraction of red giants 
show unusually strong Li lines indicative of conspicuous abundance excess.
With an aim to shed light on the origin of these peculiar stars, we 
carried out a spectroscopic study on the observational characteristics of 
selected 20 bright giants already known to be Li-rich from past studies,
in comparison with the reference sample of a large number of normal 
late G --early K giants. Our special attention was paid to clarifying any 
difference between the two samples from a comprehensive point of view (i.e., 
with respect to  stellar parameters, rotation, activity, kinematic properties, 
$^{6}$Li/$^{7}$Li ratio, and the abundances of Li, Be, C, O, Na, S, and Zn).
Our sample stars are roughly divided into ``bump/clump group'' 
and ``luminous group'' according to the positions on the HR diagram.  
Regarding the former group ($1.5 \ltsim \log (L/L_{\odot}) \ltsim 2$
and $M \sim$1.5--3~$M_{\odot}$), Li-enriched giants and 
normal giants appear practically similar in almost all respect except for Li,
suggesting that surface Li enhancement in this group may be a transient episode
which normal giants undergo at certain evolutionary stages in their lifetime.
Meanwhile, those Li-rich giants belonging to the latter group 
($\log (L/L_{\odot}) \sim 3$ and $M \sim$3--5~$M_{\odot}$) appear more 
anomalous in the sense that they tend to show higher rotation as well as 
higher activity, and that their elemental abundances (especially those 
derived from high-excitation lines) are apt to show apparent 
overabundances, though this might be due to a spurious effect reflecting
the difficulty of abundance derivation in stars of higher rotation and activity. 
Our analysis confirmed considerable Be deficiency as well as absence 
of $^{6}$Li as the general characteristics of Li-rich giants under study, 
which implies that engulfment of planets is rather unlikely for the 
origin of Li-enrichment. 
\end{abstract}

%


\section{Introduction}

While lithium (Li, $Z =3$) is an important element in the spectroscopy 
of late-type stars (e.g., for investigating physical processes of 
stellar interiors), studying its abundance is not necessarily easy 
for red giant stars (low-to-intermediate mass stars of lower surface 
temperature and larger radius, which have evolved off the main sequence 
after exhaustion of core hydrogen). That is, this element generally 
suffers considerable deficiency in the atmosphere of these evolved 
stars, due to (1) destruction of Li atoms at the hot stellar interior
of $T \gtsim 2.5\times 10^{6}$~K and (2) efficient evolution-induced 
mixing at the red-giant stage causing Li dilution in the outer envelope, 
by which the observable Li line is considerably weakened and its abundance 
determination becomes rather difficult (see e.g., Liu et al. 2014 and the 
references therein for representative Li abundance studies of red giants 
in general published so far). 
 
Interestingly, however, a small proportion ($\sim 1$\%) of red giants are 
known to show exceptionally strong Li lines, indicating a conspicuous 
overabundance of Li (by $\sim$~1--2~dex or even larger) compared to normal giants.
Since these chemically peculiar stars, called ``Li-rich giants,''
attracted interest of astrophysicists, a number of studies have been
published so far (see, e.g., Charbonnel \& Balachandran 2000; 
L\`{e}bre et al. 2009; Carlberg et al. 2010; Kumar et al. 2011;
Monaco et al. 2011; Martell \& Shetrone 2013;  Silva Aguirre et al. 2014; 
Reddy \& Lambert 2016; Casey et al. 2016; and the references therein).
Nonetheless, the reason why such significant amount of Li atoms are existent 
in their atmospheres (unlike other normal giants) has not been clarified yet, 
though various mechanisms have been proposed so far (cf. the references 
mentioned above): e.g., internal production of fresh Li atoms by the reaction of  
$^{3}{\rm He}\; (\alpha, \gamma)\; ^{7}{\rm Be}\; (e^{-}, \nu)\; ^{7}{\rm Li}$  
(so-called Cameron--Fowler mechanism) followed by an efficient transport
to the surface (for which several different processes are proposed), 
substantial supply of Li atoms by some external process such as swallowing 
of planets or brown dwarfs as a result of expansion in the red giant stage, or 
``insufficient dilution'' by which much of the original Li is still retained 
at the surface of Li-rich giants which appear superficially 
peculiar compared to other normal giants, etc.
   
In any event, much more observational work would have to be done before we 
can reasonably understand the origin of this peculiarity, given that available 
information on their characteristics is still insufficient.
Especially, comprehensive studies for a number of these peculiar giants in comparison 
with a large sample of normal giants seem to have been rarely done so far. 
That is, by observationally elucidating the similar and dissimilar aspects of 
these two groups from a diversified point of view, we may be able to find 
a clue for understanding the origin of Li peculiarity.

Conveniently, our group recently conducted extensive spectroscopic studies 
on a large number of normal red giants: Takeda, Sato, and Murata (2008; 
hereinafter referred to as Paper~I) determined the stellar parameters 
and elemental abundances of 322 late G--early K giants. Takeda and Tajitsu
(2014; hereinafter referred to as Paper~II) investigated the Be abundances
of 200 giants (sub-sample of 322 giants mentioned above) by analyzing
the Be~{\sc ii} 3131 line in the UV region. Takeda et al. (2015a; 
hereinafter referred to as Paper~III) extensively studied the abundances of 
O (along with C and Na) for 239 giants (sub-sample of 322 giants mentioned above).
Similarly, the abundances of S and Zn for these 239 giants were established
by Takeda et al. (2016b; hereinafter referred to as Paper~IV).

Given this situation, we decided to carry out a spectroscopic study on 
the observational characteristics of selected 20 Li-rich giants, based on 
high-dispersion spectra of wide wavelength coverage obtained by Subaru/HDS
in almost the same manner as done for normal giants in Papers~I--IV, 
by which we can compare the atmospheric parameters ($T_{\rm eff}$, 
$\log g$, $\cdots$), stellar parameters (e.g., $L$ or $v_{\rm e}\sin i$),
kinematic parameters (such as space velocities relative to LSR, etc.), 
and abundances of Li, Be, C, O, Na, S, and Zn between these two groups. 

The specific checkpoints we had in mind are as follows:\\
--- How are the stellar parameters of Li-rich giants compared with
those of normal giants? Is there any difference in the position
on the HR diagram between these groups?\\
--- What about the kinematic parameters characterizing the orbital motions 
in the Galaxy? Is there any difference in stellar populations?\\
--- Do they have any special characteristics regarding active stellar 
properties such as rotational velocities or magnetic activity?\\
--- We want to determine the Li abundances of Li-rich giants as precisely 
as possible. How are they compared with those of normal giants?
Are the abundance distributions both groups connected continuously or 
clearly separated?\\
--- Can we detect any signature for the existence of $^{6}$Li from the
spectra of our program stars?  Do these Li-rich giants show similar 
enrichment in Be? These two checkpoints would make an important touchstone 
to verify the planet accretion hypothesis.\\
--- Do the abundances of light elements (C, O, Na), which may be affected 
by deep mixing, show any special trend?\\
--- How about the abundance tendency of S and Zn? If Li enrichment is produced
by accretion of rocky materials, some abundance peculiarity might be observed 
in these volatile species.
 
In addition, motivated by the necessity of establishing the key properties
of the reference sample which were not addressed in our previous papers,
we newly conducted a new supplementary analysis for the large sample of normal 
giants regarding Li abundance determination and stellar activity estimation, as 
separately described in appendix~1 (activity indices of 239 field giants) and    
appendix~2 (Li abundances of 239 field giants and 103 {\it Kepler} giants).
 
\section{Observational data}

The list of our program stars (20 Li-rich giants; most of them are 
apparently bright as $V \ltsim 8$~mag) is presented in table~1, 
which were selected from the compilation of Casey et al. (2016; 
cf. table~1 therein, which they published as online material). 
All these targets have already been investigated in previous studies
(cf. table~1), of which we can make use for comparison or 
supplementary purposes. Above all, Kumar, Reddy, and Lambert's (2011) paper
is important, since 16 stars (out of 20) are included in their study, 
which may be referenced for the properties not touched in this paper 
(e.g., $^{12}$C/$^{13}$C ratio).

The spectroscopic observations of these targets were carried out by using 
the High Dispersion Spectrograph (HDS; Noguchi et al. 2002) placed at 
the Nasmyth platform of the 8.2-m Subaru Telescope on 2016 October 12--13 (UT;
two first half-nights).

On October 12, the standard Ub setting was adopted with the blue cross disperser, 
by which spectra of 13 stars (out of the 20 targets) covering 3000--4700~$\rm\AA$ 
(resolving power was $R \simeq 60000$ with the use of $0.''6$ width slit) were obtained. 
The total exposure time per star was typically $\sim$~10--20~min. In addition, 
HD~212430 was already observed with the same setting on 2013 July 17 (cf. Paper~II). 
Accordingly, UV spectra are available for 14 stars (but not for HD~174104, HD~232862,
KIC~9821622, HD~8676, PDS~100, and HD~6665). 

On October 13, the standard Ra setting was adopted with red cross disperser, 
which resulted in spectra covering 5100--7800~$\rm\AA$ (resolving power was 
$R \simeq 80000$ with the use of image slicer \#2).
The total exposure time per star was typically a few minutes to $\sim$~10~min.

The reduction of the spectra (bias subtraction, flat-fielding, 
scattered-light subtraction, spectrum extraction, wavelength calibration,
co-adding of frames to improve S/N, continuum normalization) was 
performed by using the ``echelle'' package of 
the software IRAF\footnote{IRAF is distributed
    by the National Optical Astronomy Observatories,
    which is operated by the Association of Universities for Research
    in Astronomy, Inc. under cooperative agreement with
    the National Science Foundation.} 
in a standard manner. 

Regarding the finally obtained UV spectra (October 12 observation), the typical 
S/N ratio at the Be~{\sc ii}~3131 line region is around $\sim 50$ ($\ltsim 100$). 
As to the yellow--red region spectra (October 13 observation), sufficiently high 
S/N ratios could be accomplished; i.e., regarding the Li~{\sc i} 6708 region,
S/N~$\sim$~500--1000 for most stars, except for HD~232862 ($\sim 200$),
KIC~9821622 ($\sim 90$), and PDS~100 ($\sim 250$).

\section{Atmospheric parameters}

Our determination of atmospheric parameters [effective temperature 
($T_{\rm eff}$), surface gravity ($\log g$), microturbulence ($v_{\rm t}$), 
and metallicity ([Fe/H])] was implemented in the same manner as in Paper~I
(see subsection~3.1 therein for the details) based on the equivalent widths 
($W_{\lambda}$) of Fe I and Fe II lines measured on the spectrum of each star.

Unlike the case of normal field giants studied in Paper~I, the spectral 
lines of several stars (especially HD~174104, HD~21018, HD~232862, and HD~170527) 
turned out to be appreciably rotationally-broadened. In such cases,
a special fitting function (constructed by appropriately merging the rotational- 
and Gaussian-broadening functions) was used for $W_{\lambda}$ measurement 
(if the lines are not so broadened, the conventional Gaussian-fitting
method was adopted). Even so, regarding such Li-rich giants of broad lines, 
measuring $W_{\lambda}$ was considerably difficult due to the blending effect and 
the number of available lines was small (which eventually lead to less reliable 
solutions). 

The resulting atmospheric parameters are summarized in table~1, 
while the $A$(Fe) values (Fe abundances corresponding to the final solutions) 
are plotted against $W_{\lambda}$ and $\chi_{\rm low}$ in figure~1,
where we can see that there is no systematic dependence.  
The detailed $W_{\lambda}$ and $A$(Fe) data for each star are given 
in tableE1.dat (supplementary online material).
The internal statistical errors involved with these solutions
are $\sim$~20--70~K, $\sim$~0.1--0.3~dex, $\sim$~0.1--0.4~km~s$^{-1}$,
and $\sim$~0.05--0.1~dex for $T_{\rm eff}$, $\log g$, $v_{\rm t}$, 
and [Fe/H], respectively, which tend to be somewhat larger than the case of Paper~I.
The model atmosphere for each star to be used in this study was generated
by interpolating Kurucz's (1993) ATLAS9 model grid in terms of
$T_{\rm eff}$, $\log g$, and [Fe/H]. 

The mutual correlations of these atmospheric parameters are depicted in figure~2,
where Li-rich giants (this study) and normal giants (Paper~I) are shown in
large open circles and small filled circles, respectively.
We summarize some notable characteristics read from this figure:\\
--- Generally speaking, many of our program stars occupy parameter ranges
almost consistent with those of the normal giants studied in Paper~I,
with a few exceptions as explained below.\\
--- Three stars (HD~30834, HD~205349, and HD~787) have $T_{\rm eff}$ 
as low as $\sim$~4000--4300~K, while no normal giants in Paper~I have $T_{\rm eff}$ 
lower than $\sim$~4500~K.\footnote{This is simply a matter of sample selection.
The 322 red giants studied in Paper~I were actually the targets of Okayama planet 
search program, where stars with $B-V < 1$ were preferentially selected (because 
the atmosphere of K giants with $B-V > 1$ turned out to be less stable from the
preliminary study of radial-velocity monitoring).}
In such cases, a meaningful comparison between both samples is difficult. \\
--- A similar argument holds also for HD~205349 regarding 
the surface gravity ($\log g = 0.89$), which is outside of the range 
($1.5 \ltsim \log g \ltsim 3.5$) covered by the reference sample of Paper~I.\\
--- Our program stars of Li-rich giants have metallicities
not much different from the solar value ($-0.3 \ltsim$~[Fe/H]~$\ltsim +0.3$),
while the sample giants in Paper~I include stars down to [Fe/H] $\sim -0.8$.\\  
--- One star (HD~232862) is a notable outlier, because its $\log g$ (3.79)
is unusually large for a giant. Given that this star is one of the four 
broad-line stars mentioned above, it may be possible that our parameter 
solutions for this star are not reliable. L\`{e}bre et al. (2009) adopted 
$T_{\rm eff}$ = $5000 (\pm 250)$~K, $\log g = 3.0 (\pm 0.5)$, and 
[Fe/H] = $-0.30 (\pm 0.1)$; and only $\log g$ is discrepant from ours.
We note in table~1 that $B-V$ and $T_{\rm eff}$ of this star do not conform
to the mean relation (i.e., too blue for its $T_{\rm eff}$) 
According to SIMBAD, this star is remarked as ``binary or multiple system''
and thus Hipparcos parallax is not available, which may have something to do
with this anomaly. In any event, the results for this star should be 
viewed with caution. 

The $T_{\rm eff}$, $\log g$, [Fe/H], and $v_{\rm t}$ values (along with the
luminosities and Li abundances to be addressed later in section~4 and section~7,
respectively) we derived for 20 program stars are compared with available
published values taken from various literature in figure~3. Generally speaking,
a reasonable consistency is observed for most cases, though several considerably 
discrepant cases (such as $\log g$ for HD~232862 just mentioned) are also seen.

\section{Evolutionary status}

We derived the luminosity ($L$) from the apparent magnitude, distance, 
interstellar extinction, and bolometric correction (as done in Paper~I; 
cf. subsection~3.2 therein) for 17 stars, for which Hipparcos parallaxes are available,
where we used the newly reduced data (van Leeuwen 2007).
The errors of parallaxes are on the order of $\sim 10$\% for most stars, 
which correspond to $\sim 0.1$~dex in $\log L$. Only for HD~174104, however, 
a considerably large error is exceptionally expected, since its parallax 
data given in the catalogue is $0.75 \pm 0.73$~milliarcsec.
The adopted parallax and the resulting $L$ for each star are given in table~1. 
As seen from figure~3e, our $\log L$ values tend to be slightly higher
(though not significant) than those derived by Charbonnel and Balachandran 
(2000) and Kumar et al. (2011), which may be due to their neglect of 
interstellar extinction effect. 

Regarding 3 stars (HD~232862, KIC~9821622, PDS~100) where Hipparcos data
are lacking, we tried to estimate the possible range of their $L$ values by using 
the relation $L \propto T_{\rm eff}^{4}M /g$ along with the spectroscopically 
determined $T_{\rm eff}$ and $\log g$ plotted in figure~4a, from which we can guess 
$M$ (stellar mass) by comparing them with the theoretical relations.
Since we do not have much confidence with our parameter solutions of HD~232862
(cf. section~3), we tentatively take two largely different $\log g$ values (3.8 and 3.0)
as upper and lower limits. Then, putting $M/M_{\odot} \sim 1$ and $\sim 2$ 
inferred from figure~4a for each limit. we have $0.4 \ltsim \log( L/L_{\odot}) \ltsim 1.5$.   
As to KIC~9821622 and PDS~100, we estimate from figure~4a  
$M/M_{\odot} \simeq 2 (\pm 0.5)$ and $M/M_{\odot} \simeq 3 (\pm 1)$,
which yield $1.42 < \log (L/L_{\odot}) < 1.64$ and $2.34 < \log (L/L_{\odot}) < 2.64$,
respectively. We note that such evaluated $L$ range for KIC~9821622 is consistent with
the value ($1.59 \pm 0.10$) derived by Jofr\'{e} et al. (2015), while that for 
PDS~100 is appreciably higher than Reddy et al.'s (2002) estimation (1.65)
quoted by Kumar et al. (2011).  

The resulting $\log L$ vs. $\log T_{\rm eff}$ diagram of 20 Li-rich giants stars 
are shown in figure~4b. We can see from this figure that our program stars
are roughly divided into two groups:\\
--- (i) Less luminous giants of $1.5 \ltsim \log (L/L_{\odot}) \ltsim 2$ with
$T_{\rm eff} \sim$ 4500--5000~K, presumably belonging to either RGB bump stars
or red-clump stars though these two classes are close to each other
on the HR diagram and difficult to discriminate (see, e.g., fig.~2 of 
Kumar et al. 2011).
A reasonable mass-range estimate of this group\footnote{As recently 
concluded by Takeda and Tajitsu (2015), the mass results derived in Paper~I 
are likely to be appreciably overestimated for many stars, especially for 
those in the red-clump region where tracks of different masses are 
complicatedly intersected. This is the reason why we did not derive $M$ 
for each star, unlike Paper~I. Therefore, what we can state is only a rough 
estimate regarding the possible range of $M$.} would be $M \sim$~1.5--3~$M_{\odot}$.\\
--- (ii) Luminous giants of $\log  (L/L_{\odot}) \sim 3$, 
which probably have masses around $M \sim$~3--5~$M_{\odot}$ according to the comparison 
with theoretical evolutionary tracks. We note that their $T_{\rm eff}$ values 
tend to separately cluster around $\sim 5500$~K and $\sim 4000$~K in our sample.\\ 
For convenience, we will hereinafter call these groups ``bump/clump group'' 
and ``luminous group'' respectively. Actually, this division is not only 
an apparent position-based classification but also has an intrinsic 
significance (cf. section~6 or section~9).

\section{Kinematic properties}

As done in Paper~I, we computed the kinematic parameters of 17 program stars 
(for which Hipparcos data are available) following the procedure described in 
Takeda (2007; cf. subsection~2.2 therein).
The resulting parameters (which are given in tableE2.dat of the online material) 
are compared with those of the normal giants in figure~5, where the correlations 
of $z_{\rm max}$ (maximum separation from the galactic plane) vs. $V_{\rm LSR}$ 
(rotation velocity component relative to the Local Standard of Rest: LSR), 
$e$ (orbital eccentricity) vs. $\langle R_{\rm g} \rangle$ 
(mean galactocentric radius), and $|{\bf v}_{\rm LSR}|$ 
($\equiv \sqrt{U_{\rm LSR}^{2}+V_{\rm LSR}^{2}+W_{\rm LSR}^{2}}$; amplitude of 
the velocity vector relative to LSR) vs. [Fe/H], are shown. 
The following statements can be made by inspecting figure~5:\\ 
--- Most of our Li-rich giants belong to the thin-disk population (one possible 
exception may be HD~12203, which might be of the thick-disk origin due to its
rather large $z_{\rm max}$ (0.82~kpc), just like those found by Monaco 
et al. (2011).\\
--- We can recognize that the distributions of these parameters for the Li-rich
sample and the normal sample are essentially the same, which suggests that
kinematic factors are essentially irrelevant to the origin of Li-rich phenomenon. \\
--- We observe in figure~5c that the $|{\bf v}_{\rm LSR}|$ values of our program stars 
reasonably show a decreasing tendency with [Fe/H] (as in normal giants) even 
in such a rather narrow metallicity range ($-0.3 \ltsim$~[Fe/H]~$\ltsim +0.3$).\\ 

\section{Rotation and activity}

The projected rotational velocities ($v_{\rm e}\sin i$) of 20 Li-rich giants
were evaluated in the same manner as in Paper~I  (cf. subsection~4.2 therein) 
from the macrobroadening width determined by the 6080--6089~$\rm\AA$
fitting (cf. table~2 for more information), while subtracting the contributions 
of instrumental broadening as well as $\log g$-dependent macroturbulence broadening
[cf. eq.~(1) in Paper~I]. Regarding four stars (HD~174104, HD~21018, HD~232862,
and HD~170527) where rotation is apparently dominant, we applied the rotational 
broadening function instead of the Gaussian approximation adopted in Paper~I.   
Such established spectrum fitting for each star is demonstrated in figure~6a.

We also determined the activity-sensitive index $\log R'_{\rm Kp}$ for 14 program 
stars with available UV spectra (cf. section~2), where $R'_{\rm Kp}$
is the ratio of the chromospheric emission flux of Ca~{\sc ii} K line 
at 3933.66~$\rm\AA$ (after being subtracted by the photospheric flux computed
from the relevant model atmosphere) to the total bolometric flux, following the 
procedure detailed in Takeda et al. (2012; cf. section~3 therein). The spectra in the 
core region of Ca~{\sc ii} K line, on which this measurement was made, are depicted 
in figure~6b.\footnote{The parameter ($\alpha$) controlling the integration range 
[$w_{\rm min} (\equiv 3933.7 - \alpha$), $w_{\rm max} (\equiv 3933.7 + \alpha$)] 
for evaluating the core-emission strength was chosen to be $\alpha$ = 0.9~$\rm\AA$
as the standard value, while wider values of $\alpha$ = 2.4, 1.4, and 1.9~$\rm\AA$
were adopted for HD~21018, HD~9746, and HD~205349, respectively. 
Regarding HD~170527 which shows an exceptionally strong activity among the present sample,
the integration was performed in an especially wide range of [3929~$\rm\AA$, 3938~$\rm\AA$].}   
Similarly, we also evaluated $\log R'_{\rm Kp}$ for 200 normal giants,
for which UV spectra are available (cf. Paper~II), in order to compare 
the activity nature of these two samples with each other.
This supplementary analysis is described in appendix~1.
  
The resulting $v_{\rm e}\sin i$ as well as $\log R'_{\rm Kp}$ are plotted
against $T_{\rm eff}$ in figure~6c,d (for both Li-rich and normal samples), 
and their mutual correlation is shown in figure~6e. 
We can see the following trends from these figures:\\
--- A positive correlation between $\log R'_{\rm Kp}$ and $v_{\rm e}\sin i$
is observed (figure~6e), which is reasonable since the magnetic activity 
of dynamo origin is rotation-dependent.\\
--- We see a tendency that rotation and activity tend to become lower 
with a decrease in $T_{\rm eff}$ (figure~6c,d), which is already 
known and presumably due to an enhanced magnetic braking effect 
(see, e.g., fig.~18.24 of Gray 2005).\\
--- It is evident that a large fraction of Li-rich giants have considerably 
larger $v_{\rm e}\sin i$ compared to normal giants (figure~6c,d).\\
--- However, this is not necessarily a general trend. Actually, not a few Li-rich giants 
surely follow almost the same $v_{\rm e}\sin i$ distribution (i.e., low scale
of $\ltsim 5$~km~s$^{-1}$) as the case of normal giants. Interestingly,
most of these ordinary (slower) rotators belong to the ``bump/clump'' group,
while the unusual (rapider) rotators tend to be in the ``luminous'' group
(though some exceptions do exist; e.g., HD~170527 has large $v_{\rm e}\sin i$
as well as low $\log L$). We will return to this point again in section~9.    

\section{Analysis of Li 6708 and 6104 lines}

The Li abundances of 20 program stars were determined from two line features 
of neutral lithium, the resonance line at $\sim 6707.8$~$\rm\AA$ 
(2s~$^{2}$S--2p$^{2}$P$^{\rm o}$) and the subordinate line
at 6103.6~$\rm\AA$ (2p$^{2}$P$^{\rm o}$--3d~$^{2}$D).
We employed almost the same procedure for abundance determination as adopted 
in our previous papers (e,g., Paper~III or Paper~IV), which consists of 
three consecutive steps:
(i) LTE synthetic spectrum fitting based on the numerical technique 
described in Takeda (1995) to derive the abundance of target element
as well as those of other relevant elements (cf. table~2), (ii)
inverse evaluation of the equivalent width ($W$) for the line feature
of interest (cf. table~3), and (iii) determination of LTE ($A^{\rm L}$) 
as well as non-LTE ($A^{\rm N}$) abundances from such obtained $W$, from 
which non-LTE correction $\Delta^{\rm N} (\equiv A^{\rm N} - A^{\rm L})$ can also 
be derived. We used Kurucz's (1993) WIDTH9 program for calculations of 
$A$ and $W$, which was considerably modified to enable analysis of 
multi-component lines as well as to include the non-LTE effect.   

Regarding the 6704--6711~$\rm\AA$ fitting including Li~{\sc i} 6708, 
we used the line data given in table~2 of Smith, Lambert, and Nissen (1998) 
in the 6707.2--6708.1~$\rm\AA$ region as done by Takeda and Kawanomoto (2005), 
supplemented by the data taken from the VALD database (Ryabchikova et al. 2015) 
outside of this range (though only the Fe~{\sc i} 6708.2819 line was neglected  
because VALD's $\log gf$ value for this line turned out unreasonably too large). 
Meanwhile, we adopted the data compiled in VALD for the 6101--6105~$\rm\AA$ 
fitting (including Li~{\sc i} 6104).   
In both cases, we primarily considered only lines of $^{7}$Li while neglecting $^{6}$Li 
in our analysis, because no signature was recognized for the existence of $^{6}$Li 
as mentioned below in this section. The finally accomplished spectrum 
fitting in these two wavelength regions is shown in figure~7.

The non-LTE departure coefficients necessary for deriving non-LTE abundances
were computed on the parameter grid of $T_{\rm eff}$ from 4000~K to 5500~K, 
$\log g$ from 0.0 to 4.0, $A$(Li) from 0.0 to 4.0, and [Fe/H] = 0 by following
the same procedure described in Takeda and Kawanomoto (2005), and applied to 
each star by interpolation. The trends of the non-LTE corrections
are illustrated in figure~8, where the results for 239 normal giants
(separately described in appendix~2) are also shown for the case of Li~{\sc i} 6708. 
We can see from figure~8a,b that our $\Delta^{\rm N}$ values are consistent 
with those published by Lind, Asplund, and Barklem (2009).       
We also note that the behavior of $\Delta^{\rm N}_{6708}$ for Li-rich giants
are different from those of $\Delta^{\rm N}_{6104}$ (for both Li-rich and Li-poor case) 
and $\Delta^{\rm N}_{6708}$ (for normal Li-poor case), in the sense that the former is
strongly $A$(Li)-dependent and turns from positive to negative with an increase
in $A$(Li) (figure~8e), while the latter is generally positive and 
weakly $T_{\rm eff}$-dependent (figure~8c,d). The reason for this tendency 
is explained in figure~9a, where we can see that the non-LTE effect
for the Li~6708 line acts in the direction of line-weakening (mainly due to 
the overionization effect lowering the line opacity: $l_{0}^{\rm NLTE}/l_{0}^{\rm LTE} < 1$) 
and $\Delta^{\rm N}_{6708} > 0$ for the normal Li-poor case (solid line), 
while the dilution of the line source function ($S_{l}/B <1$) begins as the abundance 
increases (dashed line) which leads to a line-strengthening and negative $\Delta^{\rm N}_{6708}$.
In contrast, such an dilution of $S_{l}$ never occurs for the weak Li~{\sc i} 6104 line
(cf. figure~9b), for which $\Delta^{\rm N}_{6104}$ is always positive. 

We first investigated whether our Li-rich giants have any $^{6}$Li in their atmospheres. 
which is important as a touchstone of ``external'' hypothesis. That is, this lighter 
isotope (which is more easily processed than $^{7}$Li and not expected to exist in normal giants) 
would be detected if swallowing of planets (presumably consisting of materials with 
$^{6}$Li/$^{7}$Li $\sim 0.1$ as observed in the interstellar matter or in meteorites) 
is the cause for Li enrichment. Although this is usually a quite delicate and difficult 
task if based on line profiles alone.\footnote{Reddy et al. (2002) and Reddy and Lambert (2016)
carried out such a detailed profile analysis of Li~{\sc i} 6708 line for Li-rich giants
(PDS~100 and HD~16771) and found that $^{6}$Li is almost absent, which is consistent 
with our consequence.} We can make use of the fact that Li abundances 
derived from the strongly saturated Li~{\sc i} 6708 line are sensitive to this isotope 
ratio while that from the weak Li~6104 line is essentially inert.
The variations of Li abundances ($\delta A$) for each line caused by increasing $^{6}$Li/$^{7}$Li 
from  0.0 to 0.1 are plotted against $W$ and $T_{\rm eff}$ in figure~10a and figure~10b, respectively. 
It is apparent from figure~10a that the extent of $\delta A_{6708}$ is appreciably large
and $W$-dependent, while $\delta A_{6104}$ is negligibly small ($\simeq 0.0$).
This sensitivity difference causes a manifest distinction in the $A^{\rm N}_{6104}$
vs. $A^{\rm N}_{6708}$ diagram shown in figure~10c ($^{6}$Li/$^{7}$Li = 0.0)
and figure~10d ($^{6}$Li/$^{7}$Li = 0.1), from which we can see that inclusion of
$^{6}$Li brings about an apparent inconsistency between the abundances derived
from two lines.\footnote{This conclusion is qualitatively the same as derived by
Balachandran et al. (2000) for HD~9746 (one of our program stars with the strongest
$W_{6708}$ of 468~m$\rm\AA$). However, their $|\delta A|$ values appear to be
quantitatively larger than ours (for both Li~{\sc i} 6708 and 6104 lines; 
cf. table~5 therein), to which the difference in the atomic line data or 
in the atmospheric parameters may be attributed.} 
Accordingly, we conclude that there is no sign of $^{6}$Li 
in our program stars. In the remainder of this paper, our discussion on the Li abundances 
is exclusively confined to those obtained with $^{6}$Li/$^{7}$Li = 0.0.

The results of the Li abundances ($A_{6708}$ and $A_{6104}$) are summarized in 
table~4, while more detailed data (including $W$ and $\Delta^{\rm N}$) are given in
tableE3.dat of the online material. Our $A^{\rm L}_{6708}$ values are mostly 
in reasonable agreement with the published results (cf. figure~3f).
The trends of $W$ and $A$(Li) for both lines in terms of $T_{\rm eff}$, [Fe/H] 
and $v_{\rm e}\sin i$ are plotted in figure~11.
The following characteristics are read from this figure.\\
--- The $W_{6708}$ values of Li-rich giants are markedly strong
($\sim$~100--500~m$\rm\AA$) compared to those of normal giants (several tens of 
m$\rm\AA$ or less) (figure~11a).\\  
--- The Li abundances of 20 Li-rich giants range from $\sim 1.9$ to $\sim 3.8$, some of 
which (e.g., HD~8676, HD~10437, HD~9746) exceed the solar-system value (3.31).\\
--- A weak trend of decreasing $A$(Li) with a decrease in $T_{\rm eff}$ is observed 
(figure~11b), which may be due to the selection effect in defining Li-rich giants 
(i.e., for the same $W$, $A$(Li) becomes smaller as $T_{\rm eff}$ is lowered). \\
--- Regarding the Li abundances of 20 program stars, no appreciable dependence
is observed upon [Fe/H] (figure~11c) as well as $v_{\rm e}\sin i$ (figure~11d),
though normal giants tend to show a weak dependence of $A$(Li) upon [Fe/H] 
as well as $v_{\rm e}\sin i$.\\  
--- Interestingly, the distribution of $A^{\rm N}_{6708}$ for normal giants 
($-1 \ltsim A({\rm Li}) \ltsim 2$) continuously connect with that of Li-rich giants 
($\sim$~1.9--3.8).

The errors in $A^{\rm N}_{6708}$ (for Li-rich giants) due to uncertainties of 
atmospheric parameters were estimated to be $\sim$~0.1--0.2~dex (for a $T_{\rm eff}$ 
change of 100~K), $\ltsim 0.02$~dex (for a $\log g$ change of 0.3~dex), 
and from $\sim$~0.01~dex [$W_{6708} \sim 100$~m$\rm\AA$] to $\sim$~0.1~dex 
[$W_{6708} \sim$~400--500~m$\rm\AA$] (for a $v_{\rm t}$ change of 0.4~km~s$^{-1}$),
which means that $T_{\rm eff}$ plays the most important role in the accuracy
of Li abundances. 

\section{Beryllium abundance}

We tried to determine the Be abundances of 14 Li-rich giant stars (for which 
UV spectra are available) by analyzing the 3130.65--3131.35~$\rm\AA$ region
comprising the Be~{\sc ii} 3131.066 line (line of $^{9}$Be, because we did not 
consider unstable $^{7}$Be in the analysis) in same manner as done in Paper~II
for 200 normal giants. However, the Be abundance solution for the spectrum fitting 
(shown in figure~12a) was obtained for only 5 stars (HD~203136, HD~212430, HD~12203, 
HD~194937, and HD~167304), while it failed to converge for the remaining 9 stars
(i.e., all what we can say about their Be abundances do not exceed 
the upper limit of $\ltsim -1$; cf. Paper~II).  
Moreover, the $W_{3131}$ values of two stars (HD~12203, HD~167304) out of these
successful 5 stars are only $\sim$~2--3~m$\rm\AA$ and almost on the same order
of the upper limit values; thus their abundances are near to the detection limit 
and unreliable (i.e., class-c in Paper~II). Regarding the representative case
of HD~194937, another theoretical spectrum corresponding to the 
solar-system Be abundance (1.42) is also overplotted in figure~12a (dashed line) 
in order to demonstrate the difference.
 
The resulting $W_{3131}$ vs. $T_{\rm eff}$, $A$(Be) vs. $T_{\rm eff}$, and
$A$(Be) vs. [Fe/H] relations for these 5 stars (as well as the those of
normal giants derived in Paper~II) are plotted in figure~12d,
figure~12e, and figure~12f, respectively.
We can see from these figures the following characteristics regarding 
the Be abundances of Li-rich giants:\\
--- As in the case of normal giants (Paper~II), photospheric Be abundances of
Li-rich giants are considerably deficient (by $\sim$~1--2~dex or more) compared 
to the solar-system abundance (1.42).\\
--- That is, there is no difference in the behavior of Be abundances 
between the Li-rich giant sample and normal giant sample, which means that
whichever mechanism causing Li-enrichment in red giants does not affect Be. \\
--- In particular, the possibility of Be overabundance is ruled out. 
This is consistent with the consequences already reported by previous
investigations (e.g., de Medeiros et al. 1997, Castilho et al. 1999,
Melo et al. 2005), though these past studies failed to determine the Be abundances 
in practice (i.e., what they could derive was only the upper limits).\\
--- This can be confirmed from figure~12b,c, where the spectra of two Li-rich giant 
are compared with those of normal Li-poor giants with similar atmospheric
parameters. We can not discern any meaningful difference in the spectrum 
feature at the position of Be~{\sc ii} lines.\\
--- This argument holds not only for $^{9}$Be~{\sc ii} lines but also for
the lines of $^{7}$Be~{\sc ii} (unstable isotope with a half-decay time of 53 days),
for which we can not recognize any signature in the spectra.
(Note that the apparent features at the wavelengths of two $^{7}$Be~{\sc ii} 3130.58 
and 3131.23 lines are mostly due to OH and Fe~{\sc i}, which are equally seen 
for Li-rich as well as normal giants.)    
It is thus unlikely that Li is being produced near to the photosphere of Li-rich 
giants by the $^{7}$Li synthesis process via $^{7}$Be (e.g., by explosive phenomenon 
such as classical novae, as recently reported by Tajitsu et al. 2015).
. 
\section{Abundances of C, O, Na, S, and Zn}

We determined the abundances of C, O, and Na for the program stars from 
C~{\sc i}~5380, [O~{\sc i}] 6300\footnote{Regarding the blending effect of 
the Ni~{\sc i} 6300.35 line on this [O~{\sc i}] line, we took its contribution 
into account by assuming [Ni/H] = [Fe/H].} as well as O~{\sc i} 7771--5, and 
Na~{\sc i} 6160 lines by following the same procedure as described in Paper~III.
Similarly, the S and Zn abundances were derived from S~{\sc i 6757} and 
Zn~{\sc i} 6362 lines in the same manner as done in Paper~IV.
The information regarding the spectrum fitting and equivalent-width
evaluation for each case is summarized in table~2 and table~3.
How the fitted theoretical spectrum matches the observed spectrum is displayed
for each star in figure~13.  
The finally obtained abundances relative to the Sun [X/H] (X = C, O, Na, S, or Zn)
are given in table~4, while more detailed data (including $W$ and $\Delta^{\rm N}$)
are presented in tableE3.dat of the online material. 
The results of $W$, $\Delta^{\rm N}$, and [X/H] for each line are plotted against 
$T_{\rm eff}$ in figure~14, where the [X/Fe] vs. [Fe/H] relation is also shown
(for both Li-rich giant and normal giant samples).

A significant tendency regarding the abundances of Li-rich giants can be noticed 
by inspection of figure~14, especially in the [X/Fe] vs. [Fe/H] plots.  
That is, we see a number of considerably discrepant points (toward direction 
of overabundance in most cases) in their [X/Fe] ratios (open circles)
derived from C~{\sc i} 5380, O~{\sc i} 7774, Na~{\sc i} 6161, and
Zn~{\sc i} 6362 in comparison to those of normal giants (filled circles), while
such a trend is not observed for [O~{\sc i}] 6300 and S~{\sc i 6757}.
Yet, we note that [X/Fe] ratios (for any element) for an appreciable fraction 
of Li-rich sample stars still show a good agreement with those of normal 
sample at the same [Fe/H].
In order to investigate the origin of this problem more in detail,
we plot the difference $\delta$[X/Fe] (deviation from the mean trend
depicted as straight lines in figure~14) against the star number 
for each star in figure~15, where the corresponding $T_{\rm eff}$, 
$\log L$, $v_{\rm e}\sin i$, and $\log R'_{\rm Kp}$ are also shown.
Interestingly, the following characteristics can be seen from figure~15:\\
--- The Li-rich giants showing conspicuously large $\delta$[X/Fe] generally 
have large $v_{\rm e}\sin i$, which are divided into higher $T_{\rm eff}$ group
($\gtsim 4800$~K$, \sim$~20--30~km~s$^{-1}$, such as star \#1, \#2, \#4, and \#7)
and lower $T_{\rm eff}$ group ($\ltsim 4500$~K, $\sim$~5--10~km~s$^{-1}$,
such as stars \#14--\#20).\\
--- In contrast, those Li-rich stars for which $\delta$[X/Fe] is fairly 
small and insignificant for any element and indiscernible from normal giants 
are slow rotators ($v_{\rm e}\sin i$ only up to a few km~s$^{-1}$; e.g., 
star \#5, \#6, \#8--\#13).\\   
--- When we examine the $\log L$ and $T_{\rm eff}$ of these two groups
(figure~15b) along with the classification mentioned in section~4, 
we see that the former (rapid rotator) group and the latter (slow rotator)
group tend to be associated with the ``luminous Li-rich giants'' and
``bump/clump Li-rich giants'', respectively. This means that those two 
groups, which were discriminated by the positions on the HR diagram, have 
different characteristics also from the viewpoint of abundance determination.

Accordingly, regarding the question ``Do Li-rich giants show any significant 
difference in the abundance trend of C, O, Na, S, or Zn compared to normal giants?'',
we may state that those Li-rich giants of the latter slowly-rotating group
are quite similar to the normal sample with respect to the abundance behavior 
of these elements. 

Yet, the situation is more complicated for the former group
of higher rotational velocities, which show apparently discrepant [X/Fe]
for several specific elements compared to normal giants. This may imply the 
existence of abundance peculiarities (mostly overabundances) also in elements 
other than Li in this group of Li-rich giants.
However, we can not exclude a possibility that this is nothing but 
a superficial effect; i.e., these abundances are not correct.
Let us focus on oxygen, for example, where [O/Fe]$_{6300}$ is almost
normal but [O/Fe]$_{7774}$ is anomalously large in these stars. 
This reminds us of the well-known trend that high-excitation O~{\sc i} 7771--5 
lines are especially strengthened in stars with high chromospheric activity 
(i.e., higher temperature in the upper layer) and the O abundance derived 
from these triplet lines by using normal radiative-equilibrium model 
atmospheres is appreciably overestimated (such as seen in Hyades K-type 
stars, e.g., section~3 in Takeda 2008 and the references therein).
Actually, recent investigation by Dupree, Avrett, and Kurucz (2016) has shown
that such a strengthening of activity-origin occurs in the O~{\sc i} triplet 
lines also for giants. If this interpretation is correct, rotation-induced 
activity might have yielded apparent overestimation of [O/Fe]$_{7774}$ 
in this group of large $v_{e}\sin i$. 
It is not clear, however, whether all the other [X/Fe] showing considerably 
large discrepancy (e.g., [C/Fe]$_{5380}$, [Na/Fe]$_{6161}$, [Zn/Fe]$_{6362}$) 
are explained by this scenario. As another possibility, inevitably increased 
abundance errors for the case of broad/shallow/blended line spectra might 
also (at least partly) be involved. 
In any case, we are not sure at present whether they represent the true abundances;
they might simply reflect the difficulty of abundance determination for such 
 Li-rich giants of high rotation and high activity.

\section{Summary and discussion}
 
We carried out a comprehensive study on various observed properties 
(stellar parameters, rotation, activity, kinematic parameters, chemical 
abundances of Li, Be, C, O, Na, S, and Zn) of 20 Li-enhanced giants and 
compared them with those of a large sample of normal red giants. 
The purpose was to clarify what is different or not different between 
these two samples. The resulting observational characteristics are summarized below. 

\begin{itemize}

\item
We found from the stellar parameters 
that our program stars are roughly divided into (i) ``bump/clump group'' 
($1.5 \ltsim \log (L/L_{\odot}) \ltsim 2$, $M \sim$~1.5--3~$M_{\odot}$) 
and (ii) ``luminous group'' ($\log  (L/L_{\odot}) \sim 3$; 
$M \sim$~3--5~$M_{\odot}$) from their positions on the HR diagram.

\item
From the viewpoint of kinematic parameters, most of our Li-rich 
sample stars belong to the thin-disk population, such as in the reference 
sample of normal red giants. This is consistent with the spectroscopically 
established metallicities, which are not much different from the solar 
metallicity ($-0.3 \ltsim$~[Fe/H]~$\ltsim +0.3$).

\item
With regard to $v_{\rm e}\sin i$ and $\log R'_{\rm Kp}$, our program stars 
of the ``luminous group'' show a tendency of rotating more rapidly (and 
more active) as compared to the sample of normal giants,
while those of the ``bump/clump group'' are quite indistinctive
in this respect (mostly slow rotators).

\item
Regarding the lithium abundances derived from Li~{\sc i} 6708 and 
Li~{\sc i} 6104 lines, we found that inclusion of $^{6}$Li ($^{6}$Li/$^{7}$Li = 0.1)
causes a systematic disagreement between these two, while consistency 
can be accomplished for the case of $^{6}$Li/$^{7}$Li = 0.0.
This suggests that $^{6}$Li is absent in the atmospheres of Li-rich giants 
in general.

\item 
As to the abundance of beryllium, the resulting $A$(Be) values are 
$A$(Be)~$\ltsim 0$ at most, indicating an appreciable deficiency as 
was the case for normal giants (Paper~II). In this sense, we do not see 
any distinct difference of $A$(Be) between the Li-rich and the normal samples; 
especially, the possibility of Be overabundance is ruled out. 

\item
Regarding the abundances of C, O, Na, S, and Zn in comparison with the sample 
of normal giants, the abundance trend of slow rotators (mostly in the 
``bump/clump group'') is practically similar and almost indiscernible.
Meanwhile, considerably discrepant abundances are seen for some 
elements (e.g., those from high-excitation lines) in Li-rich giants of 
faster rotators (mostly in the ``luminous group'), though they might be superficial 
results caused by a non-classical chromospheric effect or by increased 
difficulties of abundance determination, 
 
\end{itemize}

In concluding this paper, we make some comments on the possible Li-enrichment 
mechanism, which is still under debate, based on the observational evidences 
elucidated from this study.

Firstly, we could not recognize any signature of $^{6}$Li, and confirmed that 
Be is deficient as in normal giants, which means that any peculiarities 
in $^{6}$Li as well as in Be are not accompanied with the overabundance of Li. 
Accordingly, ``planet-engulfment hypothesis'' (accretion of unprocessed 
gas or solid material) is considered to be unlikely as the origin of
Li-rich giants (at least as a major mechanism), since it would enrich 
Be as well as $^{6}$Li.

Secondly, the fact that our program stars are roughly divided into two groups
``luminous group'' (rapid rotator of higher activity, higher-mass stars of 
$\sim$~3--5~$M_{\odot}$) and ``bump/clump group'' (slow rotator of lower activity, 
less massive stars of $M \sim$~1.5--2.5~$M_{\odot}$) may suggest that different 
origins are responsible for their Li anomaly.

Regarding the former group, many of them have larger $v_{\rm e}\sin i$ and thus 
are more active as compared to the normal giants of similar $T_{\rm eff}$, 
It seems likely that high rotation is the important key to understand 
their Li-rich phenomenon. Actually, the connection between rotation
and high Li content was already remarked by some previous studies 
(e.g., Fekel \& Balachandran 1993, Drake et al. 2002, de Laverny et al. 2003).  
Can rapid rotation itself lead to an overabundance 
of Li in the surface (e.g., enhanced mixing)?  Or large angular momentum 
is simply a by-product of some other Li-enrichment process? 

As to the latter ``bum/clump'' group, their stellar properties appear 
almost similar to those of normal giants.
That being a case, the surface Li enhancement in this group might be 
a very short-lived episode which normal giants undergo in a certain 
period in their lifetime, as has been occasionally suggested
(e.g., Charbonnel \& Balachandran 2000). Then, the question is, at which 
evolutionary stage does this phenomenon takes place: At the bump in 
the first ascent of RGB, or at the He core flash at the tip of RGB 
(cf. Kumar et al. 2011)? Maybe both are probable, since we know two 
Li-rich giants in the {\it Kepler} field with seismologically established 
evolutionary status: KIC~9821622 (locating near to the RGB bump) is 
in the H-shell burning stage (Jofr\'{e} et al. 2015), while KIC~5503307 is
a clump star in the He-burning stage (Silva Aguirre et al. 2014). 

\bigskip

This research has been carried our by using the SIMBAD database,
operated by CDS, Strasbourg, France. 

\appendix
\section{Ca~{\sc ii} 3934 core-emission strengths of 200 red giants}

In order to compare the behavior of $\log R'_{\rm Kp}$ for the 14 Li-rich giants
derived in section~6 with that of normal giants, we determined this parameter
for 200 giants (subsample of 322 stars in Paper~I), for which UV spectra used 
for Be abundance determinations in Paper~II are available. 
Since these data contain the violet region including Ca~{\sc ii} 3933.66 line, 
we derived $\log R'_{\rm Kp}$ by integrating the core emission component of 
this line in the same manner as done in section~6 by following the procedure 
detailed in Takeda et al. (2012).
The spectra of 200 stars in the relevant region are displayed in figure~16.
The resulting $\log R'_{\rm Kp}$ values are compiled in tableE4.dat of
the online material, where more detailed data (atmospheric parameters and
projected rotational velocity taken from Paper~I, integration range) are also given.
These $\log'R_{\rm Kp}$ data are shown in figure~6d,e (filled circles). 

\section{Li abundances of normal red giants}

In an attempt to compare the Li abundances of 20 Li-rich giants
(cf. section~7) with those of normal giants, we determined $A$(Li) from
Li~{\sc i} 6708 line also for the following two samples:
(1) 239 giants studied in Paper~III and Paper~IV (subsample of 322 stars 
in Paper~I), for which red-region  spectra  are available.
(2) 103 giants in the {\it Kepler} field investigated in the stellar parameter studies
by Takeda and Tajitsu (2015) and Takeda et al. (2016a), for which the evolutionary 
status is asteroseismologically established. 
The adopted procedure is essentially the same as adopted in section~7,
though the spectrum fitting was done in a slightly narrower region of
6707.0--6708.5~$\rm\AA$ (because especially broad-line stars are not included 
in this case). We could achieve satisfactory convergence for most cases, 
though the solutions failed  for the cases where the Li line feature is too 
weak and undetectable (8 out of 239 giants, 17 out of 103 {\it Kepler} giants).
The accomplished fit between theoretical and observed spectra
are shown in figure~17 (239 field giants) and figure~18 (103 {\it Kepler} giants). 
The resulting Li abundances and the related quantities
($W_{6708}$, $\Delta^{\rm N}_{6708}$, etc) and presented in tableE5.dat 
and tableE6.dat of the online material, respectively, and their trends 
are shown in figure~19. 
The $\Delta^{\rm N}_{6708}$, $W_{6708}$, and $A^{\rm N}_{6708}$(Li) values 
for the former 239 giants are also plotted (in light-blue dots) 
in figure~8 and figure~11 as the reference data of normal giants.    

Since all the 239 stars in the former sample are included in Liu et al.'s (2014) 
extensive Li abundance analysis for red giants, we can compare our results 
(LTE Li abundances and non-LTE corrections) with theirs as displayed in figure~19a,b. 
We can see from  figure~19a that our Li abundances are quite consistent 
with those of Liu et al. (2014) as far as the Li~{\sc i} 6708 line is 
sufficiently strong (group A1). However, as the line becomes weaker
(group A2), their abundances tend to be systematically larger than our results. 
The non-LTE corrections adopted by Liu et al. (2014), for which they invoked 
Lind et al.'s (2009) tables, are slightly (by a few hundredths dex) smaller 
than ours (figure~19b), though the differences are quantitatively insignificant.

Let us briefly discuss the trends of Li abundances in these (239+103) giant stars. 
Here, it should be remarked that all the converged abundance solutions are 
not necessarily reliable, because they may include fortuitously converged cases 
even if the Li line is negligibly weak. Following Takeda and Kawanomoto (2005), 
we estimate the detectability limit of $W_{6708}$ as being on the order of 
several m$\rm\AA$ which corresponds to $A$(Li)~$\sim 0$ (cf. figure~19c). 
Therefore, the abundance values of $A$(Li)$\ltsim 0$ are below the reliability 
limit (shown by a dotted line in figure~19c,d,e,f), and thus should not be seriously taken. 
We can see that Li abundance distributions of both samples (239 field giants
and 103 {\it Kepler} giants) are almost similar to each other (cf. figure~19d,e,f).  
That is, $A$(Li) shows a large dispersion ($\gtsim 2$~dex) and a weak decreasing tendency 
with a decrease in $T_{\rm eff}$ (figure~19d) as well as in $v_{\rm e}\sin i$ (figure~19f). 
This may be related to the correlation between $v_{\rm e}\sin i$ and $T_{\rm eff}$ 
(figure~19g), though the declining trend of $A$(Li) with a lowering of $T_{\rm eff}$
in {\it Kepler} RG stars (cf. figure~19d, red filled circles) may reflect 
the evolution of Li dilution during the course of ascending RGB.
No appreciable dependence upon [Fe/H] is recognized (figure~19e).

Regarding the {\it Kepler} giants (with known mass values and evolutionary status), 
$v_{\rm e}\sin i$ tends to increase with $M$ (figure~19h) and the dispersion
of $A$(Li) (and $v_{\rm e}\sin i$) is notably small at $M \sim 1.8M_{\odot}$ 
around the RC1--RC2 boundary (figure~19i), which is interesting. 
KIC~5503307, one of our {\it Kepler} sample showing a prominently strong 
Li~{\sc i} 6708 line (figure~18), is a Li-rich giant discovered by 
Silva Aguirre et al. (2014), which is classified as RC1 
(red-clump star of $M$ lower than 1.8~$M_{\odot}$ burning He in its core). 
Considering that KIC~9821622 (also studied in our main analysis) is a Li-rich 
giant known to be in the RG stage (i.e., H shell-burning star ascending 
RGB; cf. Jofr\'{e} et al. 2015), we can understand that Li-rich phenomenon 
in red giants takes place in both red-giant branch stars and red-clump stars.

\newpage
\onecolumn

\setcounter{table}{0}
\setlength{\tabcolsep}{3pt}
\begin{table}[h]
\scriptsize
\caption{Basic data and the stellar parameters of 20 Li-rich giants.}
\begin{center}
\begin{tabular}{ccccccccccccrcl}\hline\hline
\# & name & HIP  & Sp & $V$ & $B-V$ & $\pi$ & $\log L$ & $T_{\rm eff}$ & $\log g$ & $v_{\rm t}$ & [Fe/H] & $v_{\rm e}\sin i$ & $\log R'_{\rm Kp}$ & Reference\\
(1) & (2) & (3) & (4) & (5) & (6) & (7) & (8) & (9) & (10) & (11) & (12) & (13) & (14) & (15) \\
\hline
 1 & HD~174104  &   92223&  G0~Ib   &  8.36&   0.72&    0.75&   3.26& 5649&  2.05&  2.33&  $-$0.14& 28.7&$\cdots$ &  L82 \\
 2 & HD~21018   &   15807&  G5~III  &  6.37&   0.85&    2.68&   2.72& 5327&  2.05&  1.81&  +0.07& 21.2&  $-$4.70 & BN98, CB00, KRL11\\
 3 & HD~203136  &  105208&  K0      &  7.74&   0.94&    3.77&   1.98& 5084&  2.80&  1.17&  +0.14&  5.1&  $-$4.74 & CB00, KRL11\\
 4 & HD~232862  &$\cdots$&  G8~II   &  9.60&   0.73&$\cdots$& $\cdots$& 4938&  3.79&  2.02&  $-$0.20& 20.2&$\cdots$ & L09 \\
 5 & HD~212430  &  110602&  K0~III  &  5.76&   0.97&    7.11&   2.07& 4923&  2.48&  1.32&  $-$0.19&  2.0&  $-$5.00 & L14 \\
 6 & KIC~9821622&$\cdots$&  ...     &$\cdots$& $\cdots$&$\cdots$& $\cdots$& 4896&  2.91&  1.03&  $-$0.25&  1.9&$\cdots$ & J15\\
 7 & HD~170527  &   90274&  K0      &  6.84&   1.00&    6.31&   1.71& 4842&  2.57&  1.63&  $-$0.35& 22.9&  $-$3.69 & KRL11\\
 8 & HD~12203   &    9343&  G5      &  6.75&   1.00&    6.57&   1.74& 4812&  2.49&  1.33&  $-$0.32&  2.0&  $-$4.95 & KRL11\\
 9 & HD~194937  &  100953&  G9~III  &  6.23&   1.07&   10.59&   1.52& 4786&  2.56&  1.23&  $-$0.03&  1.8&  $-$5.11 & KRL11\\
10 & HD~8676    &    6647&  K0~III  &  7.77&   1.05&    4.27&   1.71& 4774&  2.53&  1.26&  $-$0.11&  1.8&$\cdots$ & KRL11\\
11 & HD~183492  &   95822&  K0~III  &  5.57&   1.05&   10.90&   1.78& 4765&  2.58&  1.25&  +0.06&  2.0&  $-$5.21 & B89, CB00, KRL11\\
12 & HD~167304  &   89246&  K0~III  &  6.36&   1.04&    6.07&   1.97& 4761&  2.43&  1.35&  +0.04&  2.8&  $-$5.21 & KRL11\\
13 & HD~10437   &    8052&  K1~III  &  6.57&   1.08&    6.74&   1.92& 4756&  2.55&  1.28&   0.00&  1.9&  $-$5.09 & KRL11\\
14 & HD~214995  &  112067&  K0~III: &  5.92&   1.11&   11.78&   1.58& 4626&  2.43&  1.25&  +0.04&  4.9&  $-$4.96 & KRL11\\
15 & PDS~100    &$\cdots$&  ...     &$\cdots$& $\cdots$&$\cdots$& $\cdots$& 4524&  1.98&  1.46&  $-$0.05&  8.3&$\cdots$ & KRL11 \\
16 & HD~6665    &    5285&  G5      &  8.44&   1.19&    4.49&   1.57& 4520&  2.61&  1.23&  +0.28&  5.2&$\cdots$ & CB00, KRL11\\
17 & HD~9746    &    7493&  K1~III: &  6.20&   1.24&    6.36&   2.11& 4490&  2.14&  1.40&  $-$0.10&  7.2&  $-$4.55 & B89, B00, CB00, KRL11\\
18 & HD~30834   &   22678&  K3~III  &  4.79&   1.41&    5.41&   2.92& 4283&  1.79&  1.45&  $-$0.24&  2.3&  $-$5.04 & B89, CB00, KRL11 \\
19 & HD~205349  &  106420&  K1~Ibvar&  6.27&   1.80&    2.45&   3.27& 4138&  0.89&  2.24&  $-$0.18&  6.5&  $-$5.38 & B89, CB00, KRL11\\
20 & HD~787     &     983&  K4~III  &  5.29&   1.48&    5.25&   2.75& 4027&  1.75&  1.31&  +0.07&  2.8&  $-$5.56 & B89, CB00, KRL11\\
\hline
\end{tabular}
\end{center}
(1) Object number (arbitrarily assigned). (2) Star name. (3) Hipparcos number. (4) Spectral type. 
(5) $V$ magnitude. (6) $B-V$ color. (7) Hipparcos parallax (in milliarcsec) (van Leeuwen 2007). 
(8) Logarithmic luminosity (in unit of $L_{\odot}$). (9) Effective temperature (in K).
(10) Logarithmic surface gravity (in cm~s$^{-2}$). (11) Microturbulent velocity dispersion (in km~s$^{-1}$).
(12) Fe abundance relative to the Sun (in dex). (13) Projected rotational velocity (in km~$^{-1}$). 
(14) Logarithmic ratio of the chromospheric Ca~{\sc ii} 3934 line core emission flux to the 
total bolometric flux (cf. Takeda et al. 2012). (15) Previous papers where the object was studied (or referred to). 
The targets are arranged in the order of decreasing $T_{\rm eff}$. The spectral type and 
the photometric data ($V$ and $B-V$) were adopted from the Hipparcos catalogue (ESA 1997), 
except for those of HD~232862 which were taken from the SIMBAD database.
Note that UV spectra (and thus $\log R'_{\rm Kp}$) are not available for six stars 
(\#1, \#4, \#6, \#10, \#15, \#16) and parallax data are lacking for three stars 
(\#4, \#6, \#15). The reference codes given in column (15) (they are by no means 
complete, because some of them are based on other studies; each paper should be consulted 
for more details) denote as follows:
L82 --- Luck (1982), B89 --- Brown et al. (1989), BN98 --- Barrado y Navacu\'{e}s et al. (1998),
CB00 --- Charbonnel and Balachandran (2000),  B00 --- Balachandran et al. (2000), 
L09 --- L\`{e}bre et al. (2009), KRL11 --- Kumar et al. (2011), L14 --- Liu et al. (2014),
J15 --- Jofr\'{e} et al. (2015).
\end{table}

\setcounter{table}{1}
\setlength{\tabcolsep}{3pt}
\begin{table}[h]
\scriptsize
\caption{Spectrum fitting analysis done in this study.}
\begin{center}
\begin{tabular}{ccccc}\hline\hline
Purpose & fitting range ($\rm\AA$) & abundances varied$^{*}$ & atomic data source & figure \\
\hline
$v_{\rm e}\sin i$ determination & 6080--6089  & Si, Ti, V, Fe, Co, Ni &KB95 (cf. Paper~I)  & figure~6a \\
Li abundance from Li~{\sc i} 6708  & 6704--6711 &  Li, Fe & SLN98+VALD (cf. section~7) & figure~7a \\
Li abundance from Li~{\sc i} 6104  & 6101--6105 &  Li, Ca, Fe & VALD (cf. section~7) & figure~7b \\
Be abundance from Be~{\sc ii} 3131 & 3130.65--3131.35 &  Be, Ti, Fe & P97 (cf. Paper~II) & figure~12a \\
C abundance from C~{\sc i} 5380   & 5378.5--5382 & C, Ti, Fe, Co & KB95 (cf. Paper~III) & figure~13a \\
O abundance from [O~{\sc i}] 6300   & 6297--6303 & O, Si, Sc, Fe & TH05+KB95 (cf. Paper~III) & figure~13b \\
O abundance from O~{\sc i} 7771--5   & 7770--7777 & O, Fe, Nd, CN & TKS98+KB95 (cf. Paper~III) & figure~13c \\
Na abundance from Na~{\sc i} 6161  & 6157--6164 & Na, Ca, Fe, Ni & KB95 (cf. Paper~III) & figure~13d \\
S abundance from S~{\sc i} 6757   & 6756.0--6758.1 & S, Fe, Co & KB95 (cf. Paper~IV) & figure~13e \\
Zn abundance from Zn~{\sc i} 6362  & 6361--6365 & O, Cr, Fe, Ni, Zn & KB95 (cf. Paper~IV) & figure~13f \\
\hline
\end{tabular}
\end{center}
$^{*}$ The abundances of other elements than these were fixed by assuming [X/H] = [Fe/H] in the fitting. 
KB95 --- Kurucz and Bell (1995), 
SLN98 --- Smith, Lambert, and Nissen (1998),
VALD --- VALD database (Ryabchikova et al. 2015),
P97 --- Primas et al. (1997),
TH05 --- Takeda and Honda (2005), and
TKS98 --- Takeda, Kawanomoto, and Sadakane (1998). 
\end{table}

\setcounter{table}{2}
\setlength{\tabcolsep}{3pt}
\begin{table}[h]
\scriptsize
\caption{Line data adopted for evaluating the equivalent widths.}
\begin{center}
\begin{tabular}{cccccl}\hline\hline
line & $W$ & $\lambda$ & $\chi_{\rm low}$ & $\log gf$ & Remark\\
     &     & ($\rm\AA$) & (eV) & (dex) & \\  
\hline
Li~{\sc i} 6708& $W_{6708}$ & 6707.756 & 0.00 & $-0.427$ & $^{7}$Li \\
 & & 6707.768 & 0.00 & $-0.206$ &  $^{7}$Li\\
 & & 6707.907 & 0.00 & $-0.932$ &  $^{7}$Li\\
 & & 6707.908 & 0.00 & $-1.161$ &  $^{7}$Li\\
 & & 6707.919 & 0.00 & $-0.712$ &  $^{7}$Li\\
 & & 6707.920 & 0.00 & $-0.932$ &  $^{7}$Li\\
 & & (6707.920) & (0.00) & ($-0.479$) & $^{6}$Li\\
 & & (6707.923) & (0.00) & ($-0.178$) & $^{6}$Li\\
 & & (6708.073) & (0.00) & ($-0.304$) & $^{6}$Li\\ 
\hline
Li~{\sc i} 6104& $W_{6104}$ & 6103.540 & 1.848 & +0.072(+0.03) & $^{7}$Li \\
 & & 6103.651 & 1.848 & +0.327(+0.03) & $^{7}$Li \\
 & & 6103.665 & 1.848 & $-0.627$(+0.03) & $^{7}$Li \\ 
 & &  (6103.574) & \multicolumn{2}{c}{(same as 6103.540)} & $^{6}$Li \\  
 & &  (6103.686) & \multicolumn{2}{c}{(same as 6103.651)} & $^{6}$Li \\  
 & &  (6103.699) & \multicolumn{2}{c}{(same as 6103.665)} & $^{6}$Li \\  
\hline
Be~{\sc ii} 3131 & $W_{3131}$ & 3131.066 & 0.000 & $-0.468$ & $^{9}$Be \\
\hline
C~{\sc i} 5380 & $W_{5380}$  & 5380.337 & 7.685 & $-1.842$ & \\
\hline
$[$O~{\sc i}$]$ 6300 & $W_{6300}$ & 6300.304 & 0.000 & $-9.717$ & \\
\hline
O~{\sc i} 7774 & $W_{7774}$ & 7774.166 & 9.146 & +0.174 & \\
\hline
Na~{\sc i} 6161 & $W_{6161}$ & 6160.747 & 2.104 & $-1.260$ & \\
\hline
S~{\sc i} 6757 & $W_{6757}$ & 6756.851 & 7.870 & $-1.760$ & \\
 & & 6757.007 & 7.870 & $-0.900$ & \\
 & & 6757.171 & 7.870 & $-0.310$ & \\
\hline
Zn~{\sc i} & $W_{6362}$ & 6362.338 & 5.796 & +0.150 & \\ 
\hline
\end{tabular}
\end{center}
See table~2 for the reference sources of these line data. Regarding the line data  
for $^{6}$Li (which was eventually neglected in our main analysis), we consulted 
SLN98 and VALD for the 6708 and 6104 lines, respectively. 
Since the $gf$ values for the $^{7}$Li~6104 lines given in VALD
are scaled according to the isotope ratio of $^{7}$Li/$^{6}$Li~$\simeq 12$ (typical
value for Interstellar matter), we recovered the true $gf$ values by adding 
a slight correction of $+0.03$~dex.
\end{table}

\setcounter{table}{3}
\setlength{\tabcolsep}{3pt}
\begin{table}[h]
\scriptsize
\caption{Resulting abundances of Li, Be, C, O, Na, S, and Zn.}
\begin{center}
\begin{tabular}{cccccrcccccc}\hline\hline
\# & name & [Fe/H] & $A_{\rm Li67}$ & $A_{\rm Li61}$ & $A_{\rm Be31}$ &
[C]$_{5380}$ & [O]$_{6300}$  & [O]$_{7774}$ & [Na]$_{6161}$ & [S]$_{6757}$ & [Zn]$_{6362}$ \\
(1) & (2) & (3) & (4) & (5) & (6) & (7) & (8) & (9) & (10) & (11) & (12) \\
\hline
 1 & HD~174104   & $-$0.14& 3.17 & 3.35 &$\cdots$& $-$0.23 &$\cdots$ &+0.31& $-$0.23& $-$0.24& +0.13\\
 2 & HD~21018    & +0.07& 2.93 & 3.29 &$\cdots$& $-$0.37 &$-$0.72 &+0.21& +0.23& $-$0.11& +0.38\\
 3 & HD~203136   & +0.14& 2.40 & 2.40 &  0.07 & $-$0.23 &$-$0.24 &+0.11& +0.36& $-$0.07& +0.06\\
 4 & HD~232862   & $-$0.20& 2.57 & 2.86 &$\cdots$& +0.43 &$\cdots$ &+0.41& $-$0.12& $\cdots$& +0.19\\
 5 & HD~212430   & $-$0.19& 1.94 & 1.80 & $-$0.63 & $-$0.38 &$-$0.27 &$-$0.16& +0.12& $-$0.19& $-$0.18\\
 6 & KIC~9821622 & $-$0.25& 1.85 & 1.67 &$\cdots$& $-$0.25 &+0.07 &$-$0.01& $-$0.18& $-$0.11& $-$0.14\\
 7 & HD~170527   & $-$0.35& 3.24 & 3.31 &$\cdots$& $-$0.27 &$-$0.42 &+0.76& $-$0.31& $-$0.68& $-$0.06\\
 8 & HD~12203    & $-$0.32& 2.14 & 2.12 & $-$1.74 & $-$0.16 &+0.02 &+0.08& $-$0.15& $-$0.13& $-$0.14\\
 9 & HD~194937   & $-$0.03& 3.15 & 3.01 & $-$0.44 & $-$0.21 &$-$0.05 &$-$0.19& +0.08& $-$0.01& $-$0.19\\
10 & HD~8676     & $-$0.11& 3.81 & 3.56 &$\cdots$& $-$0.23 &$-$0.12 &$-$0.05& $-$0.03& $-$0.14& $-$0.08\\
11 & HD~183492   & +0.06& 2.28 & 2.20 &$\cdots$& $-$0.13 &+0.04 &+0.09& +0.14& +0.05& +0.26\\
12 & HD~167304   & +0.04& 2.79 & 2.79 & $-$1.42 & $-$0.19 &+0.00 &+0.10& +0.36& +0.16& +0.12\\
13 & HD~10437    &  0.00& 3.62 & 3.44 &$\cdots$& $-$0.32 &+0.08 &+0.02& +0.11& +0.07& +0.07\\
14 & HD~214995   & +0.04& 3.06 & 3.13 &$\cdots$& +0.21 &$-$0.14 &+0.23& +0.09& +0.14& +0.39\\
15 & PDS~100     & $-$0.05& 2.59 & 2.62 &$\cdots$& +0.15 &$-$0.25 &+0.80& +0.46& $-$0.14& +0.19\\
16 & HD~6665     & +0.28& 2.80 & 2.92 &$\cdots$& +0.56 &+0.24 &+0.52& +0.47& +0.38& +0.74\\
17 & HD~9746     & $-$0.10& 3.73 & 3.90 &$\cdots$& +0.18 &$-$0.15 &+0.48& +0.03& $-$0.04& +0.00\\
18 & HD~30834    & $-$0.24& 2.63 & 2.53 &$\cdots$& +0.14 &+0.02 &$-$0.08& +0.18& $-$0.14& $-$0.31\\
19 & HD~205349   & $-$0.18& 1.88 & 2.00 &$\cdots$& +0.02 &$-$0.13 &+0.22& +0.85& $-$0.42& +0.36\\
20 & HD~787      & +0.07& 2.03 & 2.53 &$\cdots$& +1.19 &+0.22 &+0.22& +0.57& $\cdots$& +0.26\\
\hline
\end{tabular}
\end{center}
Following the star number, star name, and metallicity (Fe/H]) in Columns (1)--(3) 
(the same data as in table~1), the NLTE Li abundances derived from Li~{\sc i} 6708 as well as 
Li~{\sc i} 6104 lines (on the condition of $^{6}$Li/$^{7}$Li = 0.0) are given in Columns (4) and (5), 
and the LTE Be abundance is in Column (6).  (Note that these logarithmic number abundances 
relative to hydrogen are normalized in the usual manner as $A_{\rm H} = 12.00$.)
Columns (7)--(12) present the [X/H] values (differential abundances relative to the Sun in dex,
defined as $A^{\rm X}_{\rm star} - A^{\rm X}_{\odot}$)
for C (from C~{\sc i} 5380, NLTE), O (from [O~{\sc i}] 6300, LTE), O (from O~{\sc i} 7774, NLTE),
Na (from Na~{\sc i} 6161, NLTE), S (from S~{\sc i} 6757, NLTE), and Zn (from Zn~{\sc i} 6362),
respectively.   
\end{table}

\clearpage

\setcounter{figure}{0}
\begin{figure}[t]
\begin{center}
\includegraphics[width=12.0cm]{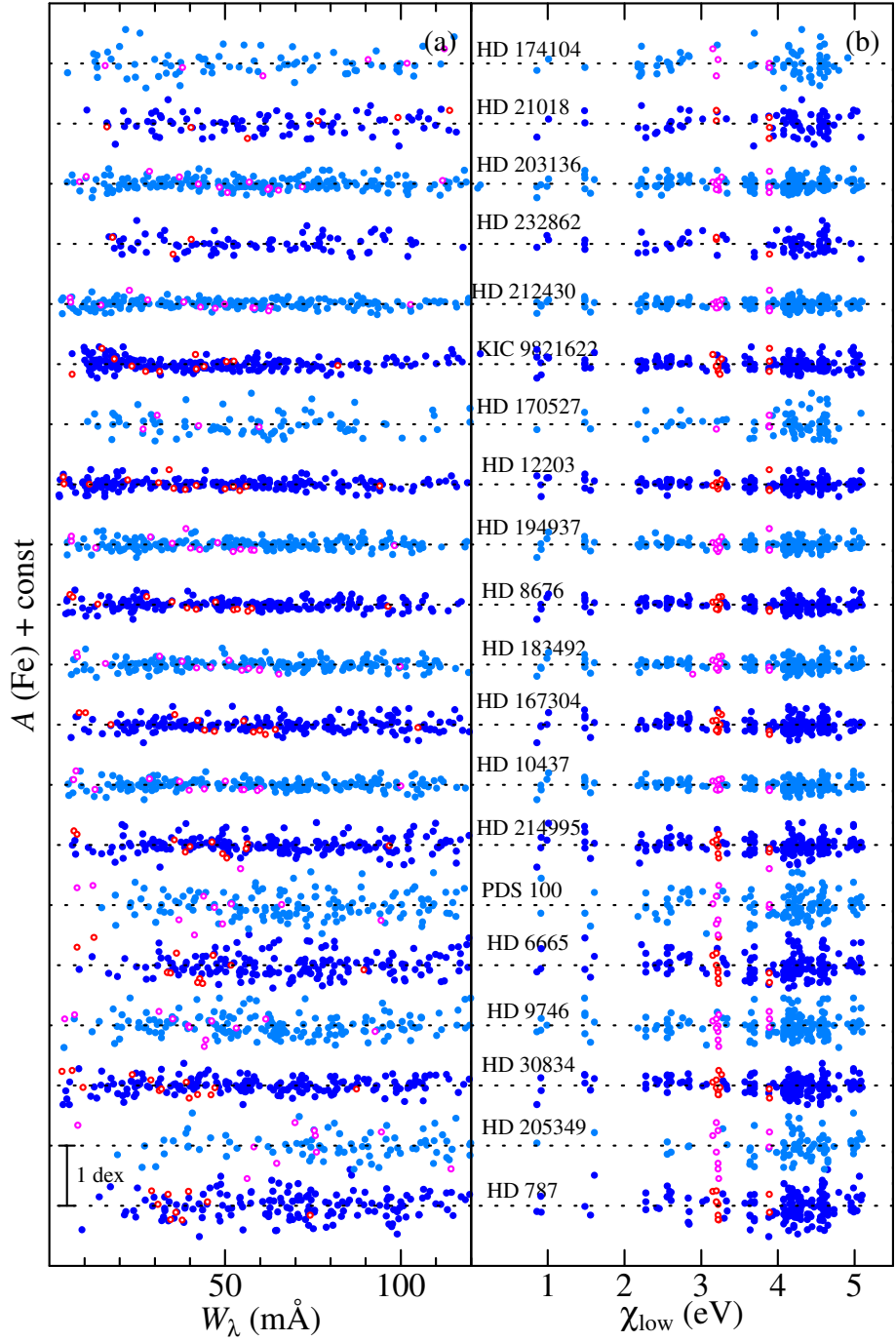}
\caption{Fe abundance vs. equivalent width relation (left panel (a))
as well as Fe abundance vs. lower excitation potential relation (right panel (b)) 
corresponding to the finally established atmospheric parameters of 
$T_{\rm eff}$, $\log g$, and $v_{\rm t}$ for each of the 20 stars,
being arranged in the decreasing order of $T_{\rm eff}$ as in table~1. 
The filled and open symbols correspond to Fe~{\sc i} and Fe~{\sc ii} 
lines, respectively. The results for each star are shown relative to 
the mean abundance (indicated by the horizontal dotted line), and 
vertically shifted by 1.0 relative to the adjacent ones.
}
\label{fig1}
\end{center}
\end{figure}

\setcounter{figure}{1}
\begin{figure}[t]
\begin{center}
\includegraphics[width=12.0cm]{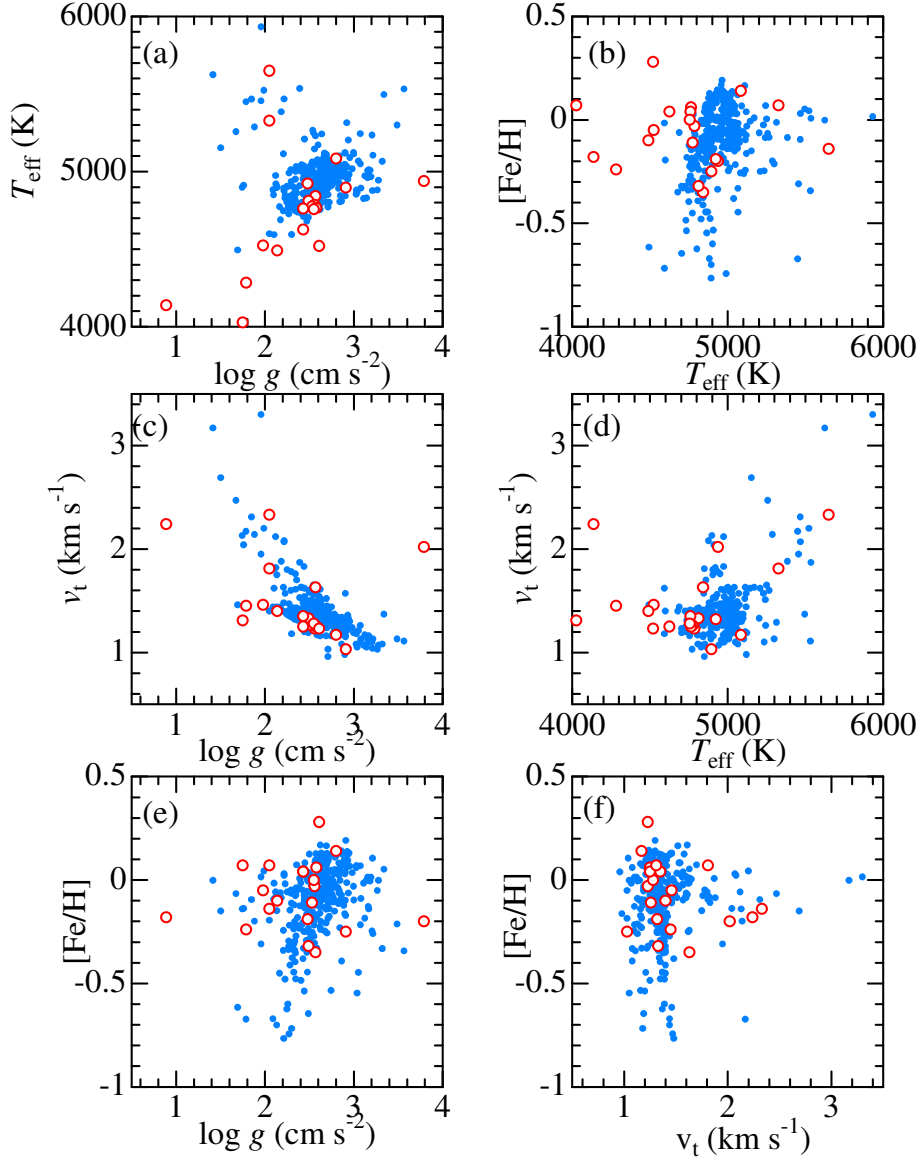}
\caption{Mutual correlations of the atmospheric parameters
spectroscopically determined by using Fe~{\sc i} and 
Fe~{\sc ii} lines for 20 Li-rich giants (this study; open circles) 
as well as 322 red giants (Paper~I; filled circles). 
(a) $T_{\rm eff}$ vs. $\log g$, (b) [Fe/H] vs. $T_{\rm eff}$,
(c) $v_{\rm t}$ vs. $\log g$, (d) $v_{\rm t}$ vs. $T_{\rm eff}$,
(e) [Fe/H] vs. $\log g$, and (f) [Fe/H] vs. $v_{\rm t}$.
}
\label{fig2}
\end{center}
\end{figure}

\setcounter{figure}{2}
\begin{figure}[t]
\begin{center}
\includegraphics[width=12.0cm]{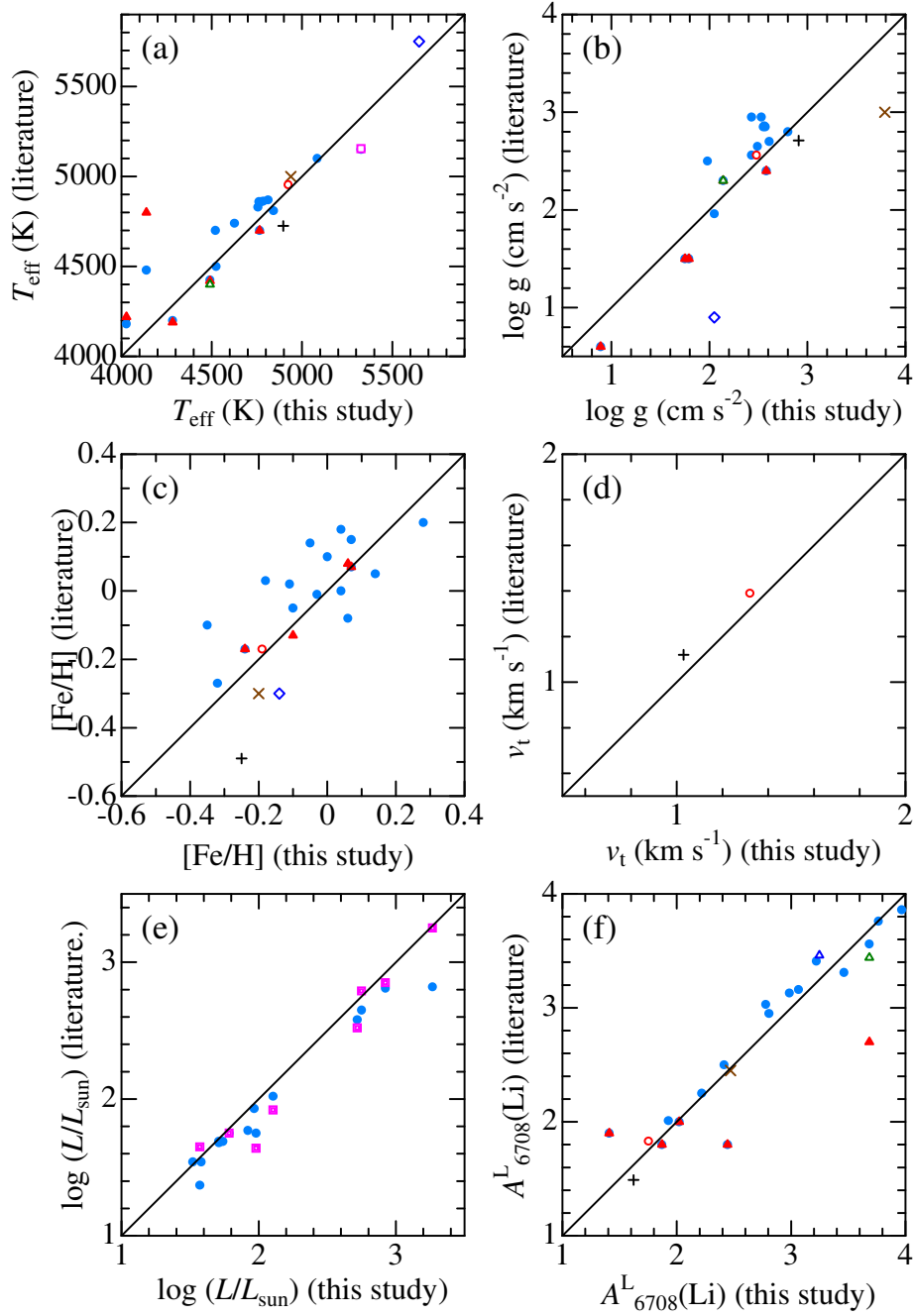}
\caption{
Comparison of the stellar parameters (and Li abundance) determined 
in this study with those reported by previous studies: 
(a) $T_{\rm eff}$, (b) $\log g$, (c) [Fe/H], (d) $v_{\rm t}$,
(e) $\log (L/L_{\odot})$, and (f) $A^{\rm L}_{6708}$(Li) 
(LTE abundance derived from Li~{\sc i} 6708).
Most of the published data were taken from table~1 of Casey et al. (2016),
though some data were newly added by ourselves (e.g., $v_{\rm t}$, $\log L$). 
Each symbol corresponds to different literature source 
(see the caption of table~1 for the meanings of the abbreviations):   
Filled circles (blue) --- KRL11,
filled triangles (red) --- B89,
open circle (red) --- L14,
open triangle (green) --- B00,
open square (pink) --- BN98,
open diamond (blue) --- L82, 
double square (pink) --- CB00, 
Greek cross (black, +) --- J15, and
St. Andrew's cross (brown, $\times$) --- L09.
Regarding the microturbulence of HD~174104, our $v_{\rm t}$ (2.3~km~s$^{-1}$) 
is largely different from L82's value (5~km~s$^{-1}$), and thus is not 
included in panel~(d).
}
\label{fig3}
\end{center}
\end{figure}

\setcounter{figure}{3}
\begin{figure}[t]
\begin{center}
\includegraphics[width=7.5cm]{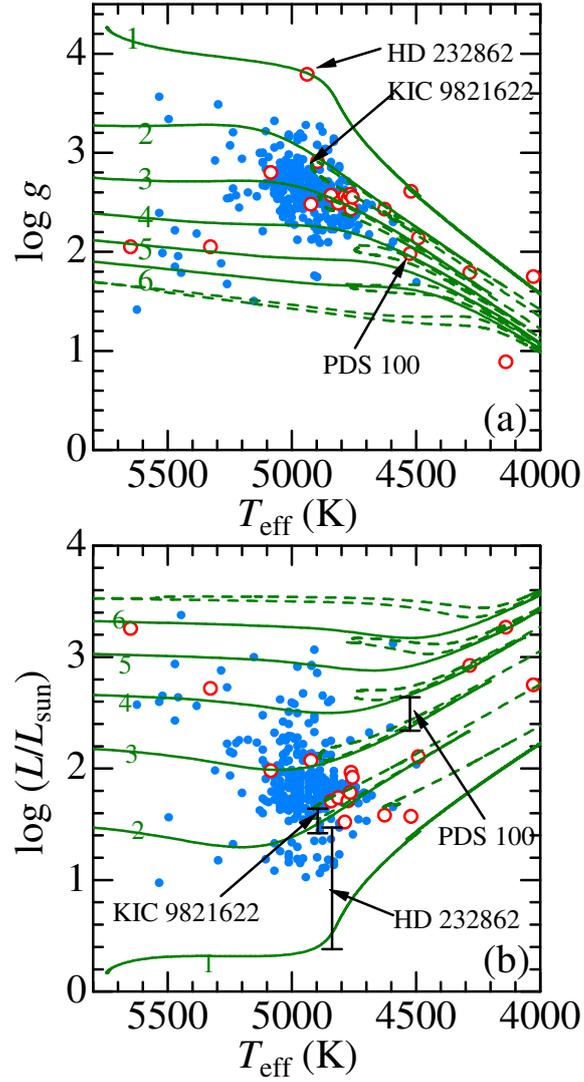}
\caption{
The $\log g$ vs. $T_{\rm eff}$ (upper panel (a)) and $\log (L/L_{\odot})$ vs. 
$T_{\rm eff}$ (lower panel (b)) diagrams based on the parameters of 20 Li-rich giants 
derived in this study (cf. table~1) and those of 322 giants taken from Paper~I,
where the meanings of the symbols are the same as in figure~2. 
The theoretical solar-metallicity evolutionary tracks for $M$ = 1, 2, 3, 4, 5, 
and 6~$M_{\odot}$ (PARSEC tracks; Bressan et al. 2012, 2013) are also depicted 
by solid lines (pre He-ignition) and dashed lines (post He-ignition) for comparison.  
See section~4 for estimations of probable $\log L$ ranges for 3 stars 
(HD~232862, KIC~9821622, and PDS~100).
}
\label{fig4}
\end{center}
\end{figure}

\setcounter{figure}{4}
\begin{figure}[t]
\begin{center}
\includegraphics[width=7.0cm]{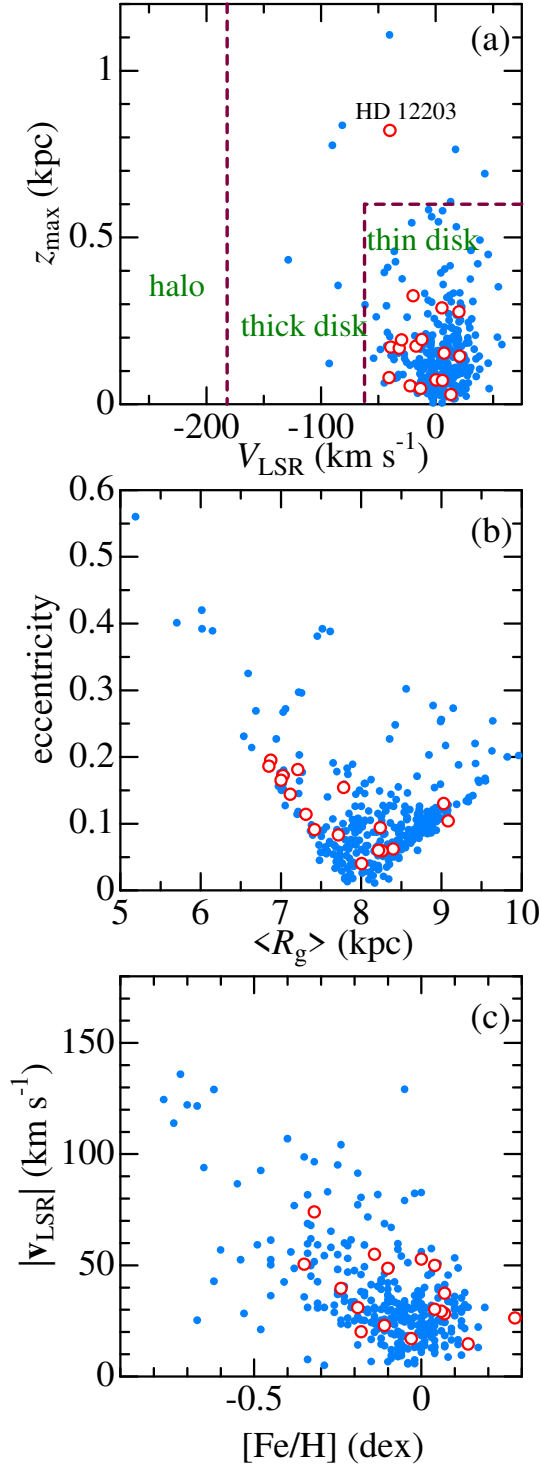}
\caption{
Mutual correlations of kinematic parameters for 17 Li-rich giants (with
available Hipparcos data) derived in this study and those of 322 giants taken 
from Paper~I (where the meanings of the symbols are the same as in figure~2):
(a) $z_{\rm max}$ (maximum separation from the galactic plane) vs. 
$V_{\rm LSR}$ (rotation velocity component relative to LSR) diagram, 
which may be used for classifying the stellar population, 
(b) $e$ (orbital eccentricity) vs. $\langle R_{\rm g} \rangle$ (mean galactocentric 
radius) diagram, and (c) [Fe/H]-dependence of the space velocity relative to LSR 
[$|{\bf v}_{\rm LSR}| \equiv (U_{\rm LSR}^{2} + V_{\rm LSR}^{2} + W_{\rm LSR}^{2})^{1/2}$].
}
\label{fig5}
\end{center}
\end{figure}

\setcounter{figure}{5}
\begin{figure}[t]
\begin{center}
\includegraphics[width=15.0cm]{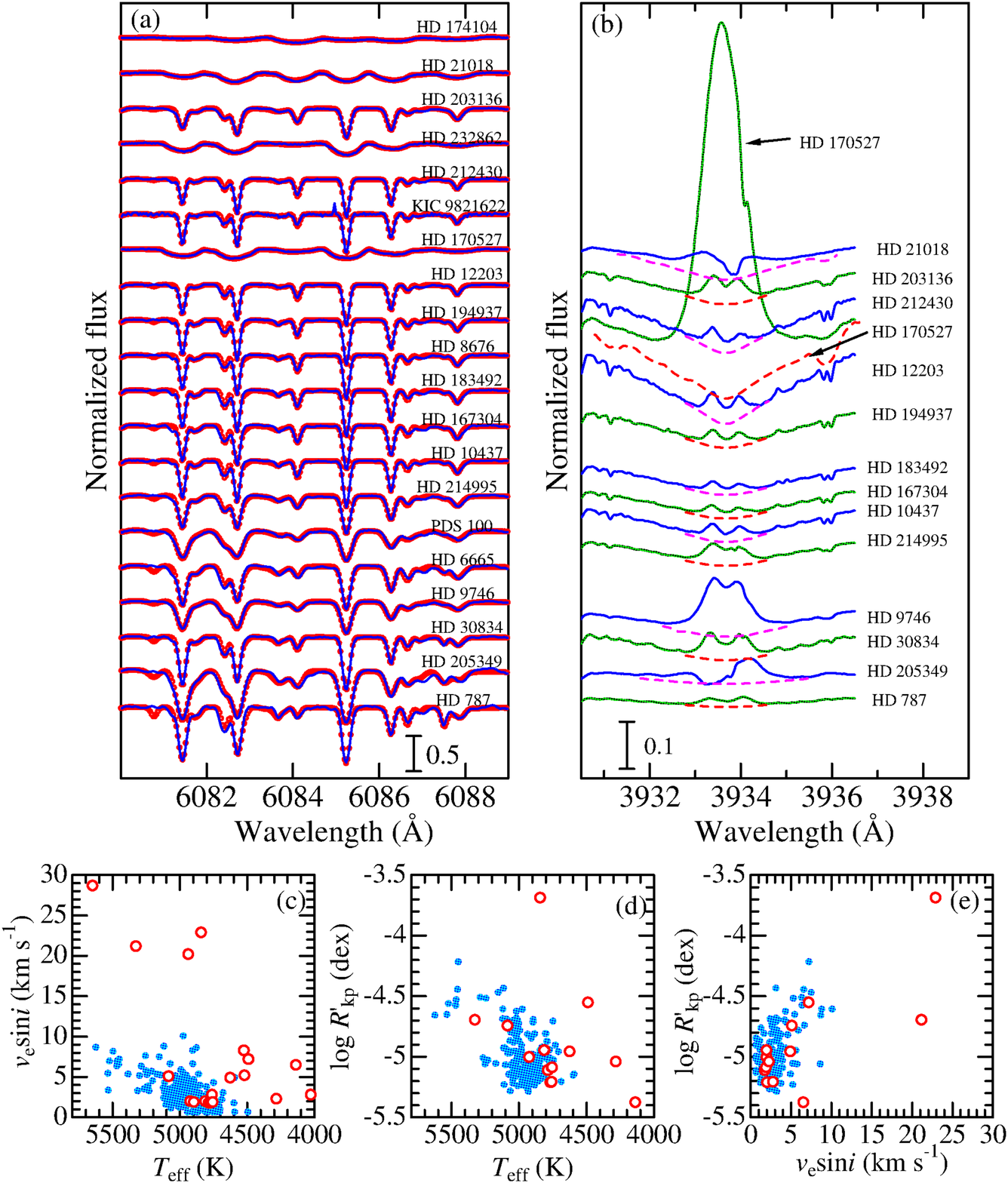}
\caption{
(a) Synthetic spectrum fitting in the 6080--6089~$\rm\AA$ 
region for evaluating $v_{\rm e}\sin i$ for 20 program stars
(see the caption of figure~7 for the meanings of the lines and symbols).
(b) Display of the 3930--3937~$\rm\AA$ region spectra (solid lines) 
including Ca~{\sc ii} K line at 3933.663~$\rm\AA$ for 17 Li-rich giants,  
for which UV spectra are available. The theoretical line profiles 
calculated from LTE photospheric model atmospheres are also overplotted 
by dashed lines in the core region [$w_{\rm min}$, $w_{\rm max}$] 
where integration for evaluating the core-emission strength was performed.
The wavelength scale is adjusted to the laboratory frame. 
Panels (c)--(e) show the mutual correlations between $\log R'_{\rm Kp}$
(activity indicator), $v_{\rm e}\sin i$, and $T_{\rm eff}$, where 
the meanings of the symbols are the same as in figure~2.
Note that [number of Li-rich sample,  number of normal sample] is
[20, 322] in panel~(c), while [14, 200] in panels (d) and (e) (because
$\log R'_{\rm Kp}$ is involved).  
}
\label{fig6}
\end{center}
\end{figure}

\setcounter{figure}{6}
\begin{figure}[t]
\begin{center}
\includegraphics[width=12.0cm]{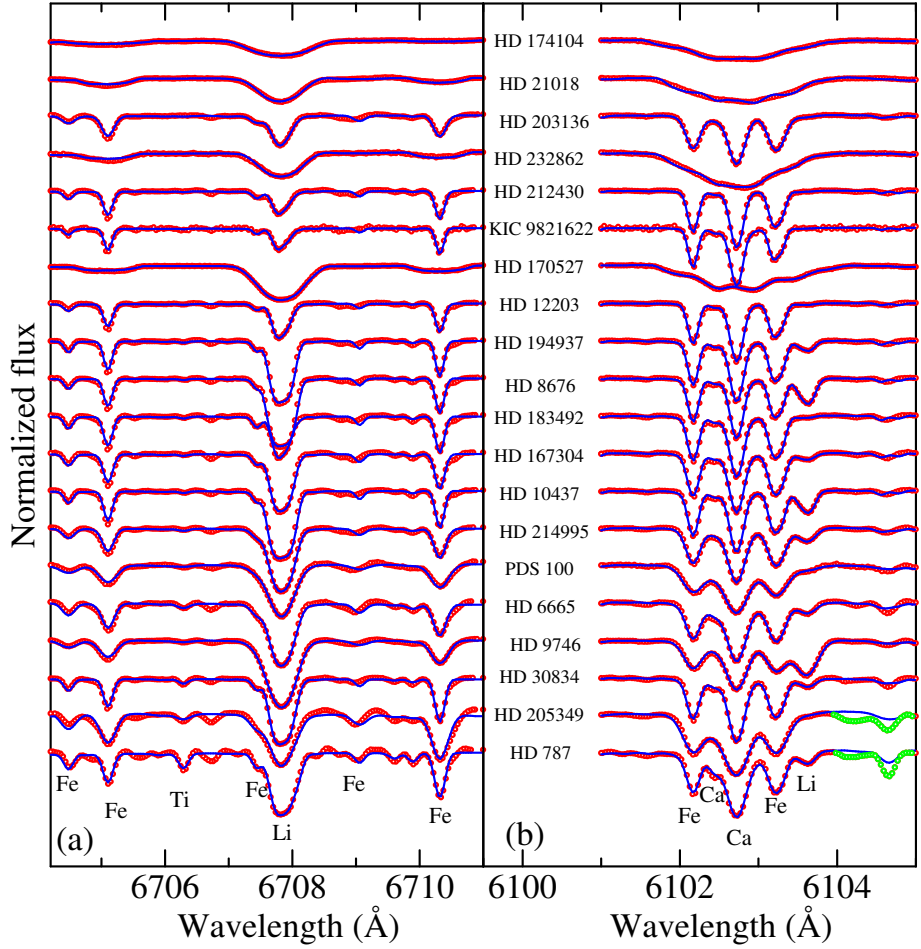}
\caption{Synthetic spectrum fitting in the 6704--6711~$\rm\AA$ 
region comprising Li~{\sc i} 6708 line (left panel~(a)) and 
6101--6105~$\rm\AA$ region comprising Li~{\sc i} 6104 line (right panel~(b)) 
for determining the Li abundances of 20 program stars
(arranged in the descending order of $T_{\rm eff}$ as in table~1).
The observed spectra are plotted in red symbols (where the masked regions 
discarded in judging the goodness of fit are colored in light-green)
while the best-fit theoretical spectra are shown in blue solid lines. 
The wavelength scale is adjusted to the laboratory system.  
}
\label{fig7}
\end{center}
\end{figure}

\setcounter{figure}{7}
\begin{figure}[t]
\begin{center}
\includegraphics[width=12.0cm]{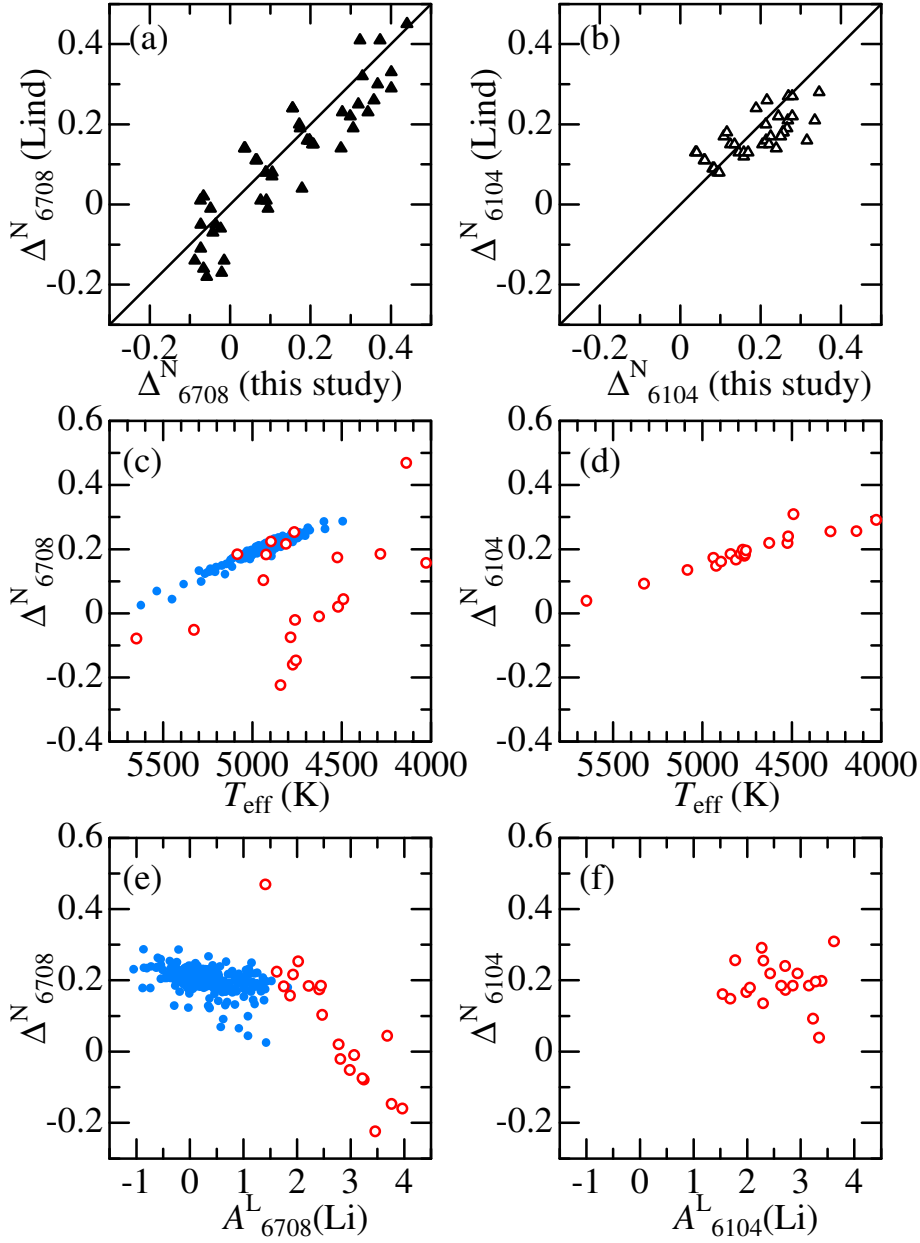}
\caption{
The top panels (a, b) compare our non-LTE corrections of Li lines ($\Delta^{\rm N}$) 
computed for 48 models ($v_{\rm t}$ = 2~km~s$^{-1}$ case with combinations of $T_{\rm eff}$ 
= 4000, 4500, 5000~K; $\log g$ = 1, 2, 3, 4; and $A^{\rm L}$(Li) = 1.5, 3.0) 
with those published by Lind et al. (2009). In the middle (c, d) and bottom (e, f) panels 
are plotted our $\Delta^{\rm N}$ results derived for actual stars against $T_{\rm eff}$
and $A^{\rm L}$(Li) (LTE abundance), respectively. The left-hand and right-hand
panels correspond to the results for Li~{\sc i} 6708 line and Li~{\sc i} 6104 line,
respectively. Regarding the four panels (c)--(f), the same meanings of the 
symbols as in figure~2.
}
\label{fig8}
\end{center}
\end{figure}

\setcounter{figure}{8}
\begin{figure}[t]
\begin{center}
\includegraphics[width=8.0cm]{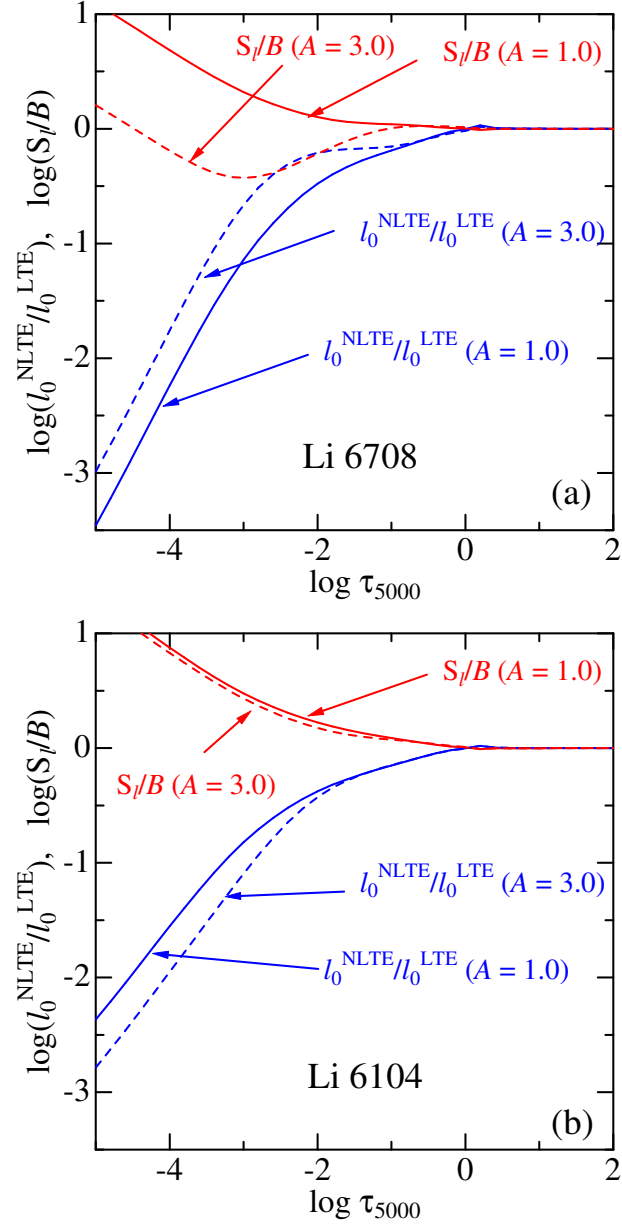}
\caption{
Ratio of the Li~{\sc i} line source function ($S_{\rm L}$) to the local
Planck function ($B$) and the NLTE-to-LTE line-center opacity
ratio as functions of the standard continuum optical depth at 5000 $\rm\AA$, 
which were computed for the solar-metallicity model of $T_{\rm eff}$ = 4500~K 
and $\log g= 2.0$ with two different Li abundances of $A$(Li) = 1 (dashed line) 
and 3 (solid line).
The red and blue lines correspond to $S_{\rm L}/B$ and 
$l_{0}^{\rm NLTE}/l_{0}^{\rm LTE}$, respectively.
Upper panel (a): 2s~$^{2}{\rm S}$ -- 2p~$^{2}{\rm P}^{\rm o}$
transition of  multiplet 1 (corresponding to Li~{\sc i} 6708). 
Lower panel (b): 2p~$^{2}{\rm P}^{\rm o}$ -- 3d~$^{2}{\rm D}$
transition of multiplet 4 (corresponding to Li~{\sc i} 6104).
}
\label{fig9}
\end{center}
\end{figure}

\setcounter{figure}{9}
\begin{figure}[t]
\begin{center}
\includegraphics[width=12.0cm]{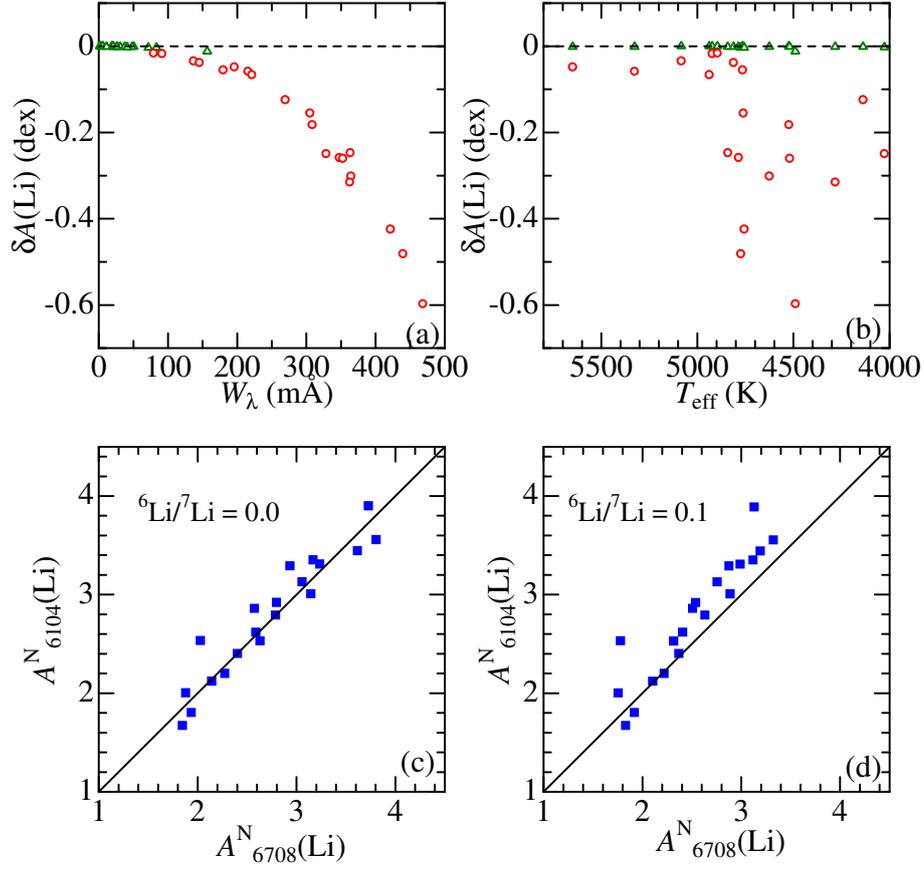}
\caption{In the upper panels (a) and (b) are plotted $\delta A$(Li) 
against $W_{\lambda}$ and $T_{\rm eff}$, respectively, where $\delta A$(Li)
is the (non-LTE) Li abundance variation caused by including $^{6}$Li with 
the isotope ratio of $^{6}$Li/$^{7}$Li = 0.1 in comparison to the standard 
case of neglecting $^{6}$Li. The open circles and open triangles
correspond to Li~{\sc i} 6708 and Li~{\sc i}~6104, respectively.
The lower two  panels show how $A^{\rm N}_{6104}$(Li) vs.
$A^{\rm N}_{6708}$(Li) correlation is affected by inclusion of $^{6}$Li.
(c) --- $^{6}$Li/$^{7}$Li = 0.0 (no $^{6}$Li), 
(d) --- $^{6}$Li/$^{7}$Li = 0.1.
}
\label{fig10}
\end{center}
\end{figure}

\setcounter{figure}{10}
\begin{figure}[t]
\begin{center}
\includegraphics[width=12.0cm]{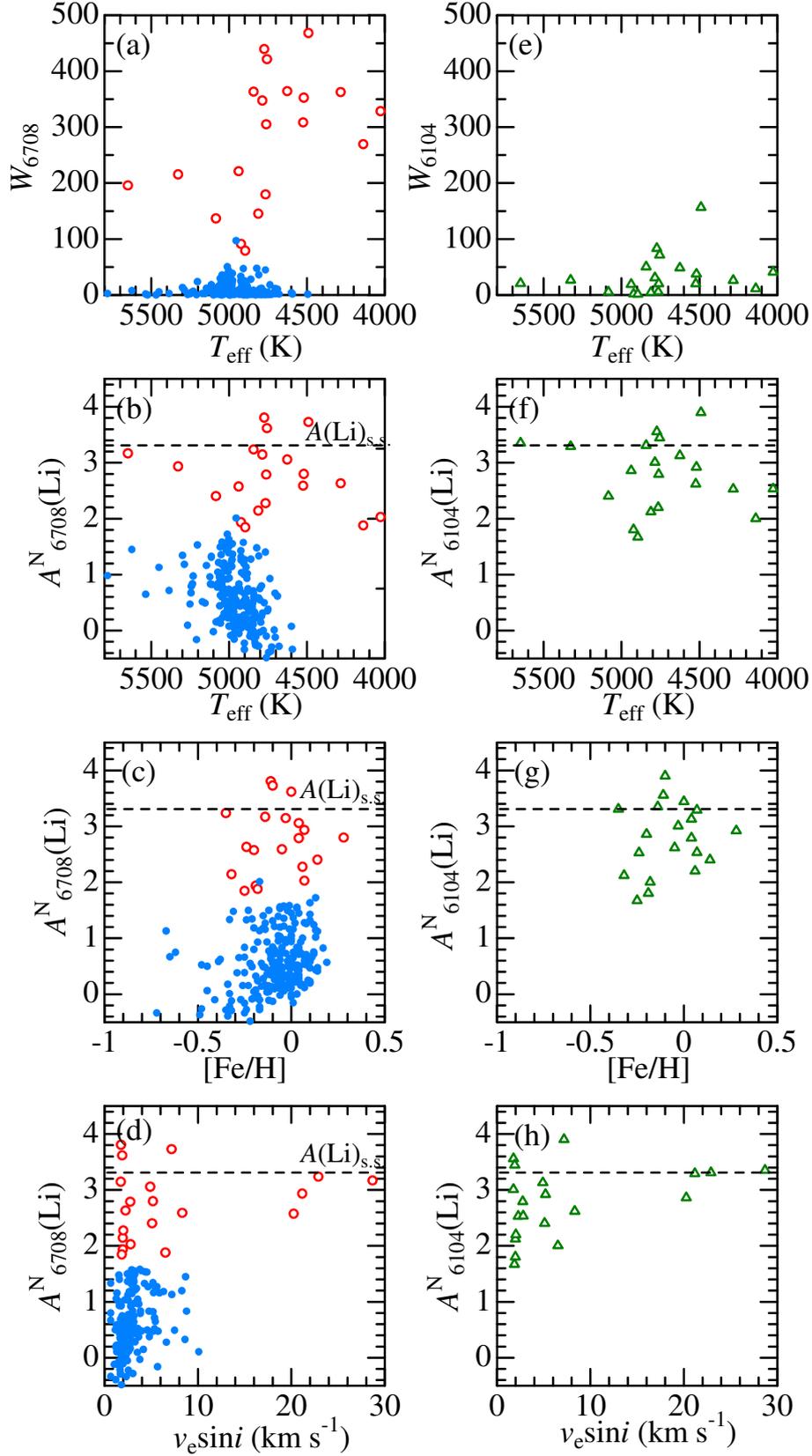}
\caption{Behaviors of $W$ (equivalent width of Li line feature) and $A^{\rm N}$(Li)
(non-LTE Li abundance) in terms of the stellar parameters. The left-hand panels 
show the results for the Li~{\sc i} 6708 line: (a) $W_{6708}$ vs. $T_{\rm eff}$,
(b)  $A^{\rm N}_{6708}$(Li) vs. $T_{\rm eff}$, (c) $A^{\rm N}_{6708}$(Li) vs. [Fe/H],
and (d) $A^{\rm N}_{6708}$(Li) vs. $v_{\rm e}\sin i$. The right hand panels (e)--(h)
(each corresponding to panels (a)--(d), respectively) are for the cases of the 
Li~{\sc i} 6104 line.  The horizontally drawn dashed line indicates the solar-system
Li abundance (3.31). The results for 20 Li-rich giants and those for 239 normal giants 
(cf. appendix~2) are shown by open and filled symbols, respectively.}
\label{fig11}
\end{center}
\end{figure}

\setcounter{figure}{11}
\begin{figure}[t]
\begin{center}
\includegraphics[width=15cm]{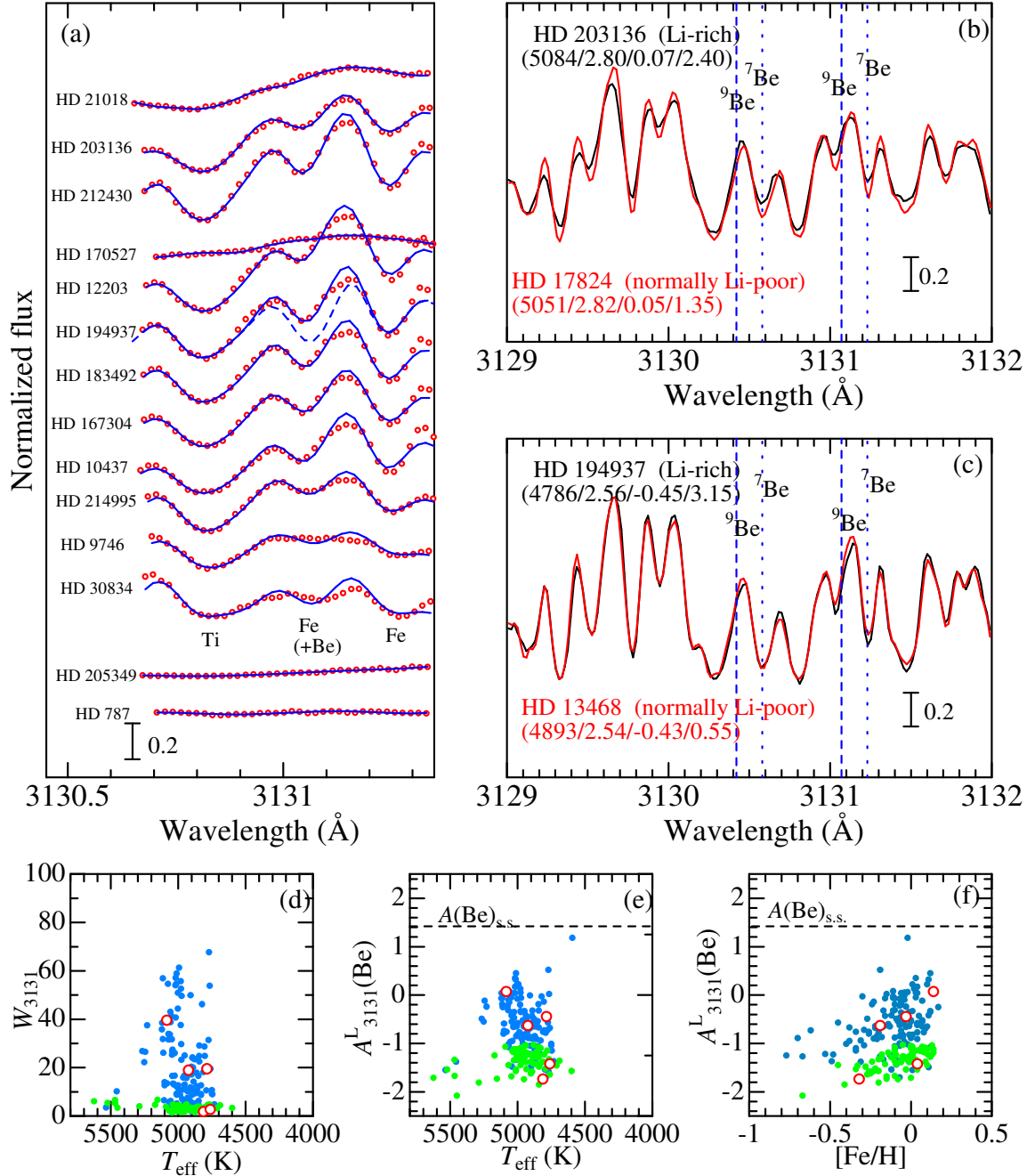}
\caption{
(a) Synthetic spectrum fitting in the 3130.65--3131.35~$\rm\AA$ region for Be abundance 
determination.  Note that although the $A$(Be) solution was converged only for 5 stars 
(HD~203136, HD~212430, HD~12203, HD~194937, and HD~167304), the fitting was anyway
accomplished also for unsuccessful cases by fixing $A$(Be) at a negligible level.
For HD~194937, an additional theoretical spectrum simulated for $A$(Be) = 1.42 
(solar-system Be abundance) is also depicted in dashed line, in order to show 
the difference compared with the best-fit spectrum of $A$(Be) = $-0.44$ (solid line). 
Otherwise, the same as in figure~7. In panels (b) and (c) are compared
the 3129--3132~$\rm\AA$ region spectra of two Li-rich giants (HD~203136 and HD~194937;
black lines) with those of normal giants of similar parameters (HD~17824 and HD~13468;
red lines), respectively. [$T_{\rm eff}$/$\log g$/$A^{\rm L}$(Be)/$A^{\rm N}$(Li) are 
indicated for each star in the figure, and the positions of stable $^{9}$Be as well as
unstable $^{7}$Be doublet lines are indicated by vertical dashed and dotted lines,
respectively.] Panels (d), (e), and (f) show the $W_{3131}$ vs. $T_{\rm eff}$, 
$A^{\rm L}$(Be) vs. $T_{\rm eff}$, and $A^{\rm L}$(Be) vs. [Fe/H] relations, respectively, 
where the meanings of the symbols are almost the same as in figure~2 (but the 
filled circles for normal giants colored in light-green indicate the upper limits) 
and the horizontal dashed line indicates the solar-system Be abundance of 1.42.
}
\label{fig12}
\end{center}
\end{figure}

\setcounter{figure}{12}
\begin{figure}[t]
\begin{center}
\includegraphics[width=16cm]{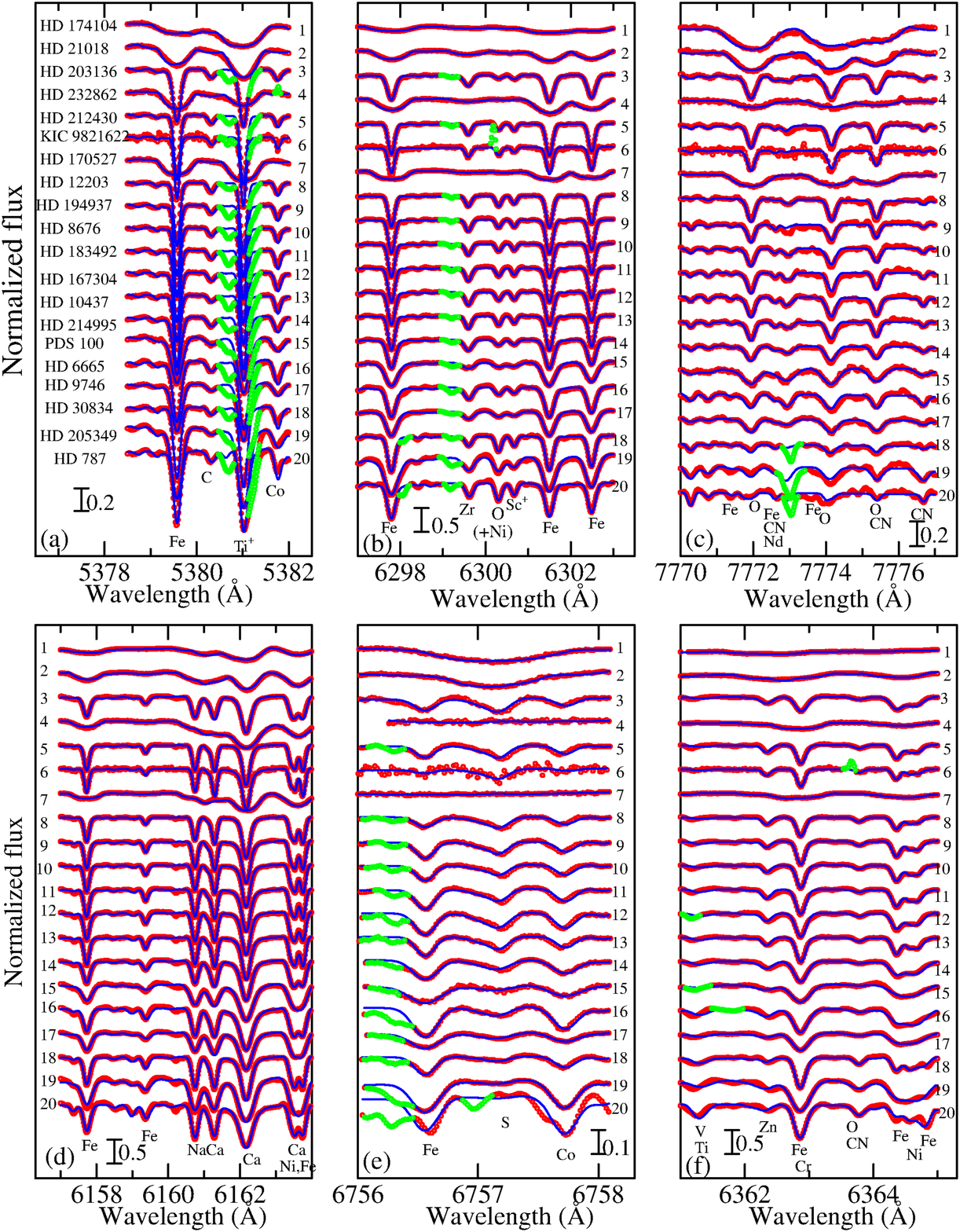}
\caption{
Synthetic spectrum fitting carried out for determining the abundances of 
C, O, Na, S, and Zn. (cf. table~2). 
(a) 5378.5--5382~$\rm\AA$ region comprising C~{\sc i} 5380, 
(b) 6297--6303~$\rm\AA$ region comprising [O~{\sc i}] 6300, 
(c) 7770--7777~$\rm\AA$ region comprising O~{\sc i} 7771--5,
(d) 6157--6164~$\rm\AA$ region comprising Na~{\sc i} 6161,
(e) 6756.0--6758.1~$\rm\AA$ region comprising S~{\sc i} 6757, and
(f) 6361--6365~$\rm\AA$ region comprising Zn~{\sc i} 6362.
Otherwise, the same as in figure~7.
Note that the telluric lines in the 6297--6303~$\rm\AA$ region and the 
broad Ca~{\sc i} autoionization feature in the 6361--6365~$\rm\AA$ region
were removed by dividing the spectra by appropriate references.
}
\label{fig13}
\end{center}
\end{figure}

\setcounter{figure}{13}
\begin{figure}[t]
\begin{center}
\includegraphics[width=14cm]{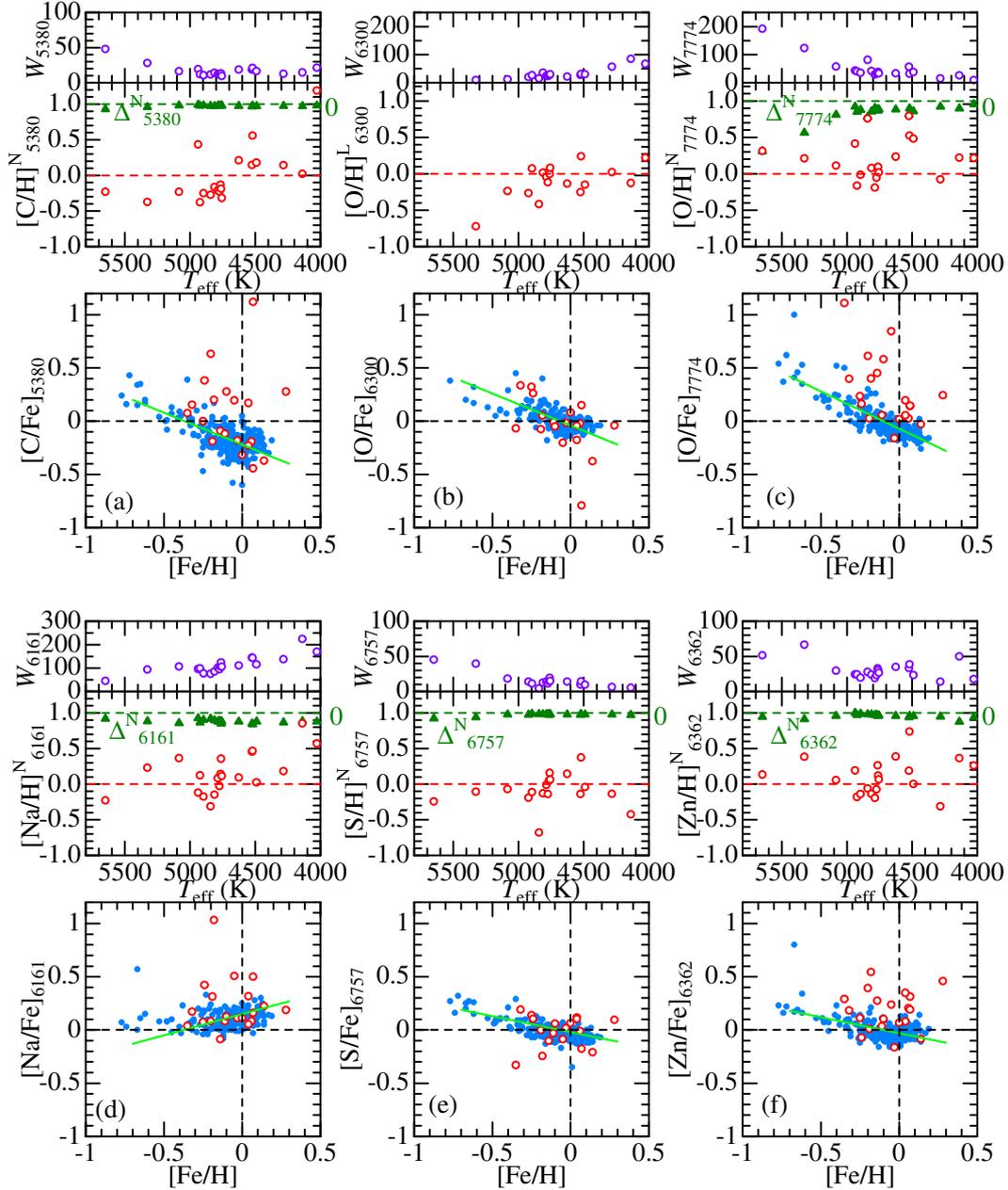}
\caption{
Each of the 6 figure sets (a--f) consist of 3 panels (tentatively named t, m, b
as abbreviations of ``top'', ``middle'', and ``bottom''), where
the results of C, O, Na, S, and Zn abundances ([X/H] $\cdots$ panel-m, 
[X/Fe] $\cdots$ panel-b), equivalent widths ($W$ $\cdots$ panel-t), 
and non-LTE corrections ($\Delta^{\rm N}$ $\cdots$ panel-m) plotted against $T_{\rm eff}$ 
(panel-t and panel-m) or [Fe/H] (panel-b).  Here, as in figure~2, the open and 
filled circles correspond to Li-rich giants and normal giants, respectively. 
(a) C~{\sc i} 5380, (b) [O~{\sc i}] 6300, (c) O~{\sc i} 7774, (d) Na~{\sc i} 6161, 
(e) S~{\sc i} 6757, and Zn~{\sc i} 6362. Note that the origin of $\Delta^{\rm N}$
(shown by filled triangles) is marked in the right ordinate of panel-m.
The (light-green) straight line drawn in panel-b is the approximate mean relation
for the normal giants (filled circles). 
}
\label{fig14}
\end{center}
\end{figure}

\setcounter{figure}{14}
\begin{figure}[t]
\begin{center}
\includegraphics[width=8cm]{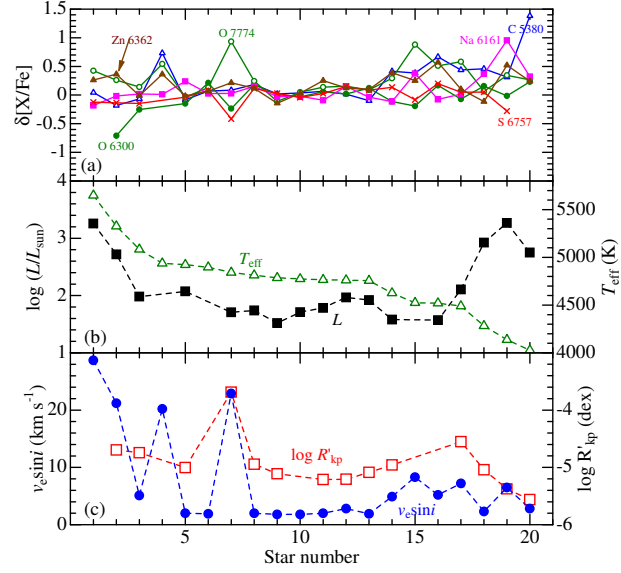}
\caption{
(a) Deviations of $\delta$[X/Fe] ($\equiv$ [X/Fe] $-$ [X/Fe]$_{\rm mean}$) 
are plotted against the star number (in the order of decreasing $T_{\rm eff}$).
Open triangles (blue) --- C~{\sc i} 5380,
filled circles (green) --- [O~{\sc i}] 6300,
open circles (green) --- O~{\sc } 7774,
filled squares (pink) --- Na~{\sc i} 6161,
crosses (red) --- S~{\sc i} 6757, and
filled triangles (brown) --- Zn~{\sc i} 6362.
(b) $\log (L/L_{\odot})$ (filled squares) and $T_{\rm eff}$ (open triangles) 
plotted against the star number.
(c) $v_{\rm e}\sin i$ (filled circles) and $\log R'_{\rm Kp}$ (open squares)
plotted against the star number.
}
\label{fig15}
\end{center}
\end{figure}

\setcounter{figure}{15}
\begin{figure}[t]
\begin{center}
\includegraphics[width=15.0cm]{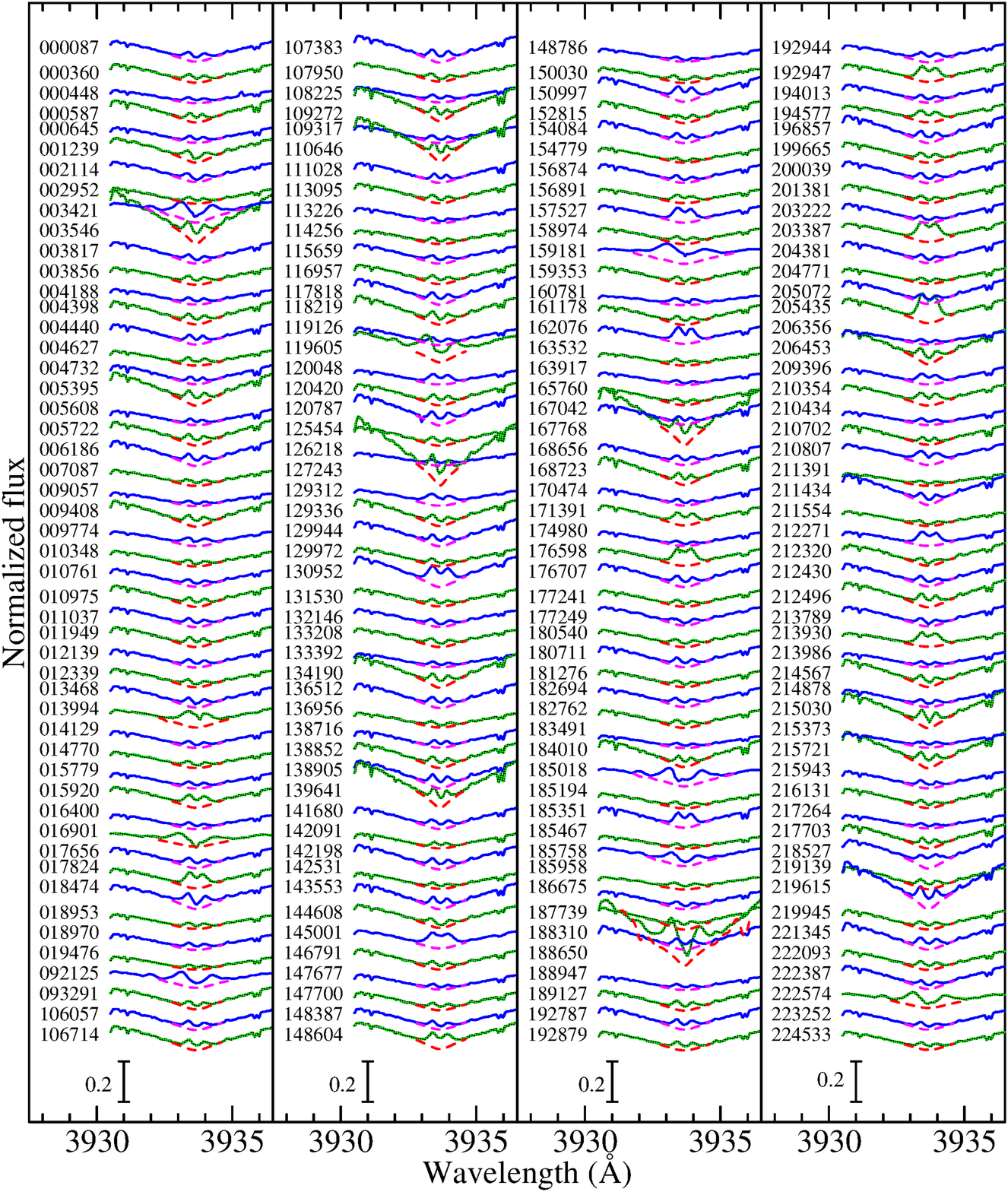}
\caption{
Display of the 3930.5--3936.5~$\rm\AA$ region spectra (solid lines) of 
Ca~{\sc ii} K line at 3933.663~$\rm\AA$ for all the 200 giants studied in Paper~II. 
In addition, theoretical LTE line profiles  ($r_{\lambda}^{\rm th}$)
calculated from the model atmospheres are also overplotted by dashed lines
in the core region [$w_{\rm min}$, $w_{\rm max}$], where integration was done 
for evaluating the emission strength. 
Each spectrum is vertically shifted by 0.1 (in continuum unit)
relative to the adjacent one. The wavelength scale of all stellar spectra 
is adjusted to the laboratory frame by correcting the radial velocity
shifts. The HD numbers are indicated in the figure. 
Regarding several lower-metallicity stars, more sharply V-shaped wings tend to
cross the spectra of other stars, which makes their identification confusing. 
In such cases, it is recommended to pay attention (not to the line wing but) 
to the line center of the theoretical profile (dashed line).
}
\label{fig16}
\end{center}
\end{figure}

\setcounter{figure}{16}
\begin{figure}[t]
\begin{center}
\includegraphics[width=14.0cm]{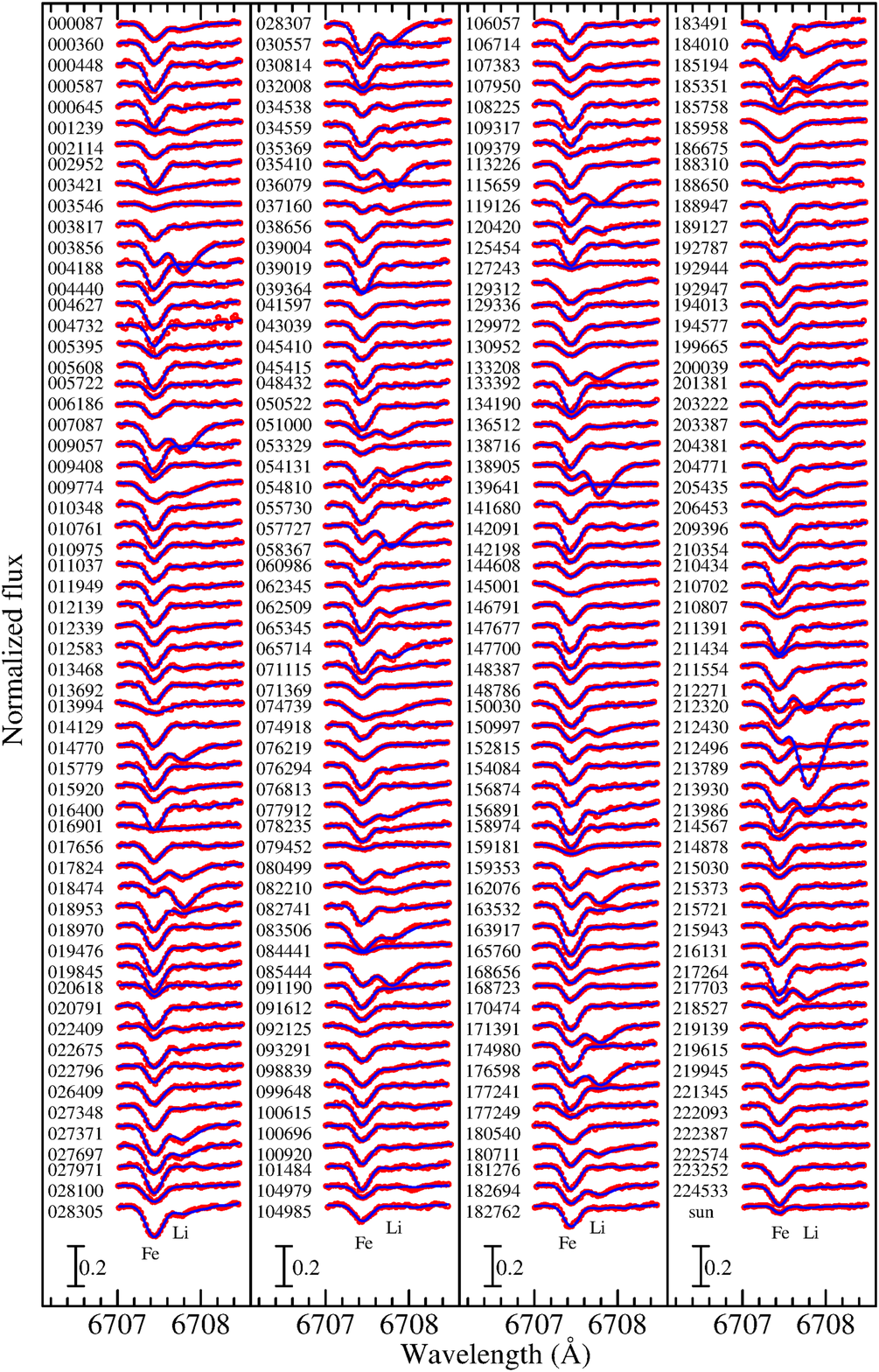}
\caption{
Synthetic spectrum analysis of the 6707.0--6708.5~$\rm\AA$ region
including lines of Li~{\sc i} and Fe~{\sc i} for 239 giants studied in Paper~III
(note that only HD~212430 is in common with our sample of 20 Li-rich giants).
The best-fit theoretical spectra are shown by solid lines, while the observed
data are plotted by symbols. A vertical offset of 0.1 is applied to each spectra 
relative to the adjacent ones. The HD numbers are indicated in the figure.
Note that fitting was somehow accomplished even for unsuccessful cases
(i.e., Li abundance solution could not be established because the line is too weak) 
by fixing the contribution of Li~{\sc i} 6708 line at a negligible level.
}
\label{fig17}
\end{center}
\end{figure}

\setcounter{figure}{17}
\begin{figure}[t]
\begin{center}
\includegraphics[width=12.0cm]{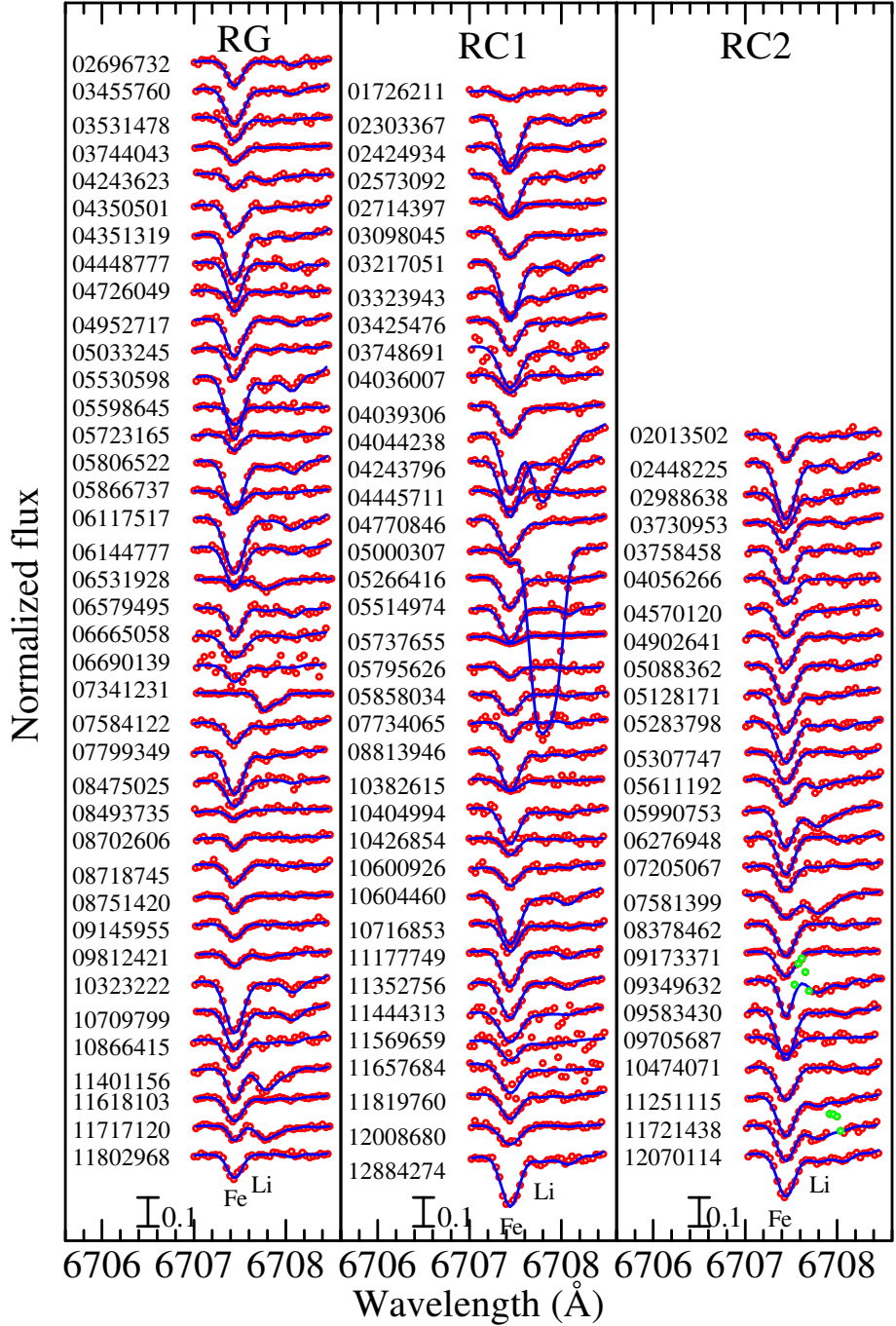}
\caption{
Synthetic spectrum analysis of the 6707.0--6708.5~$\rm\AA$ region for 103 giants 
in the {\it Kepler} field studied by Takeda and Tajitsu (2015) and Takeda et al. (2016a),
for which the evolutionary status is asteroseismologically established.
The left, center, and right panel corresponds to RG (red giants ascending
the RGB), RC1 (1st clump giants), RC2 (2nd clump giants), respectively.   
The KIC numbers are indicated in the figure. Otherwise, the same as in figure~17.
}
\label{fig18}
\end{center}
\end{figure}

\setcounter{figure}{18}
\begin{figure}[t]
\begin{center}
\includegraphics[width=14cm]{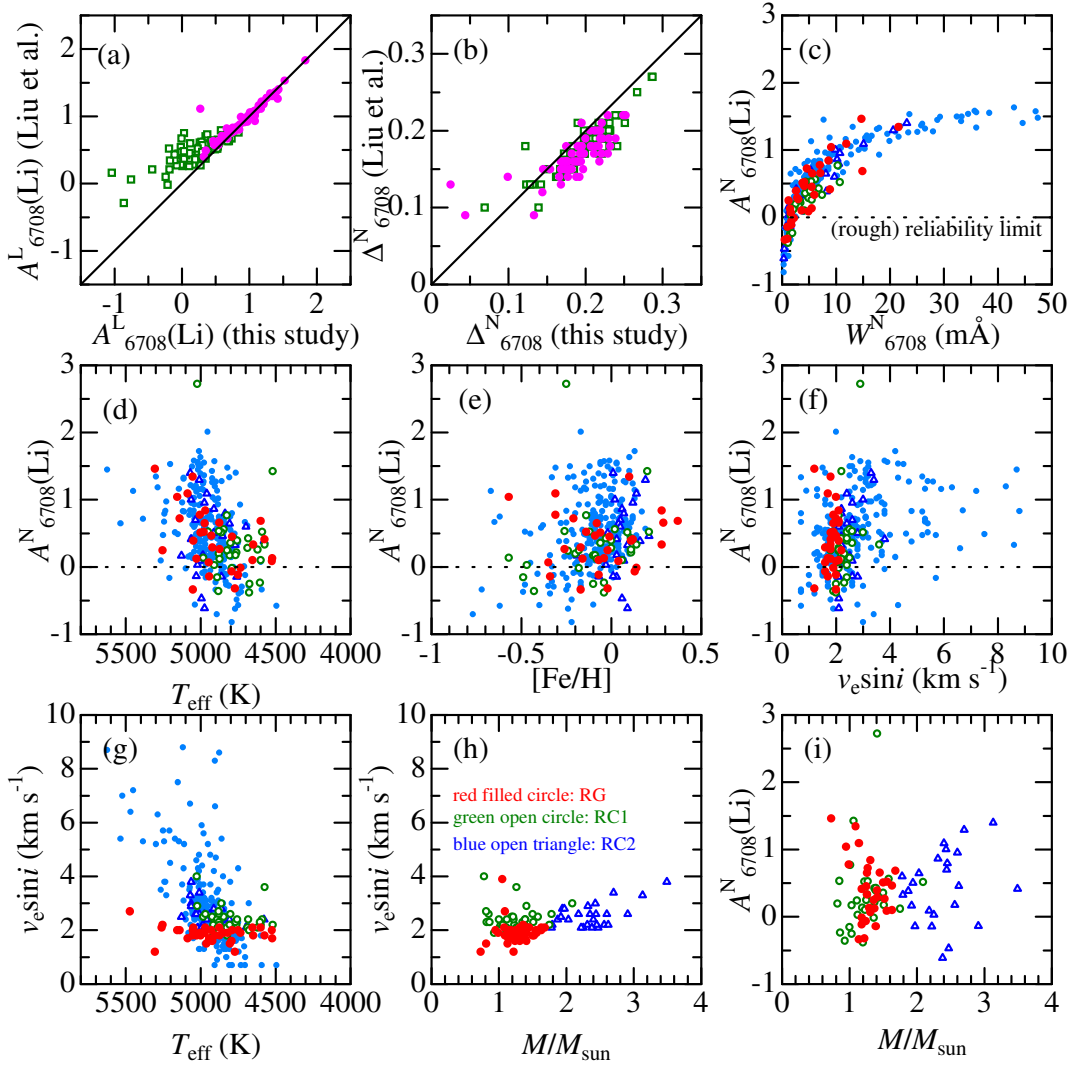}
\caption{
Panels (a) and (b) compare our lithium abundances and the corresponding 
non-LTE corrections determined for 239 giants with those derived by 
Liu et al. (2014), where filled and open symbols correspond to the subsample 
group A1 (clear Li detection) and A2 (undistinguished Li detection)
defined by them, respectively. Panels (c)--(i) illustrate the trend of 
Li abundances (along with the equivalent widths and stellar parameters) of 
239 giants (blue smaller filled circles) and 103 {\it Kepler} field giants 
(red larger filled circles --- RG, green open circles --- RC1, and blue open triangles --- RC2):
(c) $A^{\rm N}_{6708}$ vs. $W_{6708}$, (d) $A^{\rm N}_{6708}$ vs. $T_{\rm eff}$,
(e) $A^{\rm N}_{6708}$ vs. [Fe/H], (f) $A^{\rm N}_{6708}$ vs. $v_{\rm e}\sin i$,
(g) $v_{\rm e}\sin i$ vs. $T_{\rm eff}$, (h) $v_{\rm e}\sin i$ vs. $M$,
and (i) $A^{\rm N}_{6708}$ vs.$M$. Note that the lithium abundances are
reliable only above the  critical limit of $A$(Li)$\sim 0$ (shown by the
dotted line, corresponding to $W_{6708}$ of a few m$\rm\AA$), below which the 
apparent $A$(Li) values should not be seriously taken.
}
\label{fig19}
\end{center}
\end{figure}

\end{document}